\documentclass[floatfix,aps,twocolumn,10pt,superscriptaddress,prapplied]{revtex4-2}

\usepackage{bm}
\usepackage{graphicx}
\usepackage{subfigure}
\usepackage{wrapfig}
\usepackage{epstopdf}
\usepackage{siunitx}
\usepackage{braket}
\usepackage{tabularx}
\usepackage{blindtext}
\usepackage{amsmath}
\usepackage{float}
\usepackage{tabularx}
\usepackage{multirow}

\usepackage{hyperref}

\usepackage{color}
\usepackage[section]{placeins}

\def\bem#1{\begin{mathletters}\label{#1}}
\def\eml{\end{mathletters}}

\def\ket#1{{|#1\rangle}}

\def\4#1{{\boldsymbol{#1}}}
\def\8#1{{\widetilde{#1}}}

\begin{document}

\title {Enhanced quantum properties of shallow diamond atomic defects through nitrogen surface termination}

\author{R. Malkinson} 
\affiliation{Dept. of Applied Physics, Rachel and Selim School of Engineering, Hebrew University, Jerusalem 91904, Israel}

\author{M. K. Kuntumalla} 
\affiliation{Schulich Faculty of Chemistry, Technion – Israel Institute of Technology, Haifa 32000, Israel}

\author{A. Hoffman}
\affiliation{Schulich Faculty of Chemistry, Technion – Israel Institute of Technology, Haifa 32000, Israel}

\author{N. Bar-Gill}
\affiliation{Dept. of Applied Physics, Rachel and Selim School of Engineering, Hebrew University, Jerusalem 91904, Israel}
\affiliation{The Racah Institute of Physics, The Hebrew University of Jerusalem, Jerusalem 91904, Israel}
\affiliation{The Center for Nanoscience and Nanotechnology, The Hebrew University of Jerusalem, Jerusalem 91904, Israel}

\begin{abstract}
  Nitrogen vacancy (NV) centers in diamond have emerged in recent years as leading quantum sensors in various modalities. Most applications benefit from shallow NVs, enabling higher sensitivity and resolution. However, near surface NVs ($<$ 20 nm depth) suffer from reduced stability and coherence properties due to additional noise. We demonstrate a novel surface termination technique based on nitrogen plasma under non-damaging conditions, achieving significant improvement in NV optical stability and quantum coherence.
\end{abstract}

\maketitle

\section{Introduction}
Recent years have seen significant interest in optically active atomic defects in the solid-state in the context of various quantum science and technology applications, notably in quantum sensing \cite{degen2017review, taylor2008magnetometer, acosta2009diamonds, doherty2013nitrogen}. One of the leading realizations of such a platform is the nitrogen vacancy (NV) center in diamond, which exhibits favorable spin properties even at room temperature \cite{doherty2013nitrogen, bargill2013coherence, childress2006coherent}. 

For many sensing applications, it is desirable to optimize samples with shallow defects, such that their sensitivity to signals of interest outside of their host material will be maximized. However, it has been identified both theoretically and experimentally that various noise sources associated with the surface have an adverse effect on the useful properties of these shallow defects \cite{romach2015spectroscopy, stacey2019sp2, myers2017surfaceT1, kim2015decoherence, ofori2012spin, chou2017nitrogen}.

Specifically for NV centers in diamond, several approaches for mitigating adverse surface effects have been suggested and demonstrated \cite{chou2017nitrogen, kawai2019nitrogen, bluvstein2019surfaceCharge, bluvstein2019extendingSurface, sangtawesin2019origins}. Nevertheless, a robust and well-controlled general technique for diamond surface termination is still of significant interest for a broad community.

In this paper, we build upon previous proposals and demonstrations \cite{chou2017nitrogen, kawai2019nitrogen, kageura2017effect}, yet develop and implement extremely well-controlled ultrahigh vacuum non-damaging diamond surface nitridation. We combine this with in-situ spectroscopic characterization and quantum spin measurements, to demonstrate a clear advance compared to the state-of-the-art in shallow NV ($< 5$ nm from the surface) fluorescence stability and quantum coherence, reaching on average $T_2 > 4 \mu$s.

\section{Results}
Our analysis was carried out on a pure diamond sample grown by chemical vapor deposition (CVD, an "electronic grade" sample from Element Six), which was implanted with $^{15}$N ions at an energy of 2 keV and a dose of $5 \times 10^{10}$ $[\frac{ions}{cm^2}]$ (Innovion), creating a low density layer (optically addressable single NVs) of shallow NVs (approx. 3-5 nm from the surface based on SRIM calculations).

The sample was then acid cleaned (see Methods for details), and surface-terminated using both damaging and non-damaging $^{14}$N RF Plasma.

We compare the surface characteristics of a non-terminated surface, to surfaces terminated with N plasma under non-damaging and damaging conditions. This comparison addresses the chemical and structural properties of the diamond surface, as well as the quantum properties of shallow NV centers.

\subsection{Chemical and structural analysis}
\begin{figure}[tbh]
{\includegraphics[width = 0.9 \linewidth]{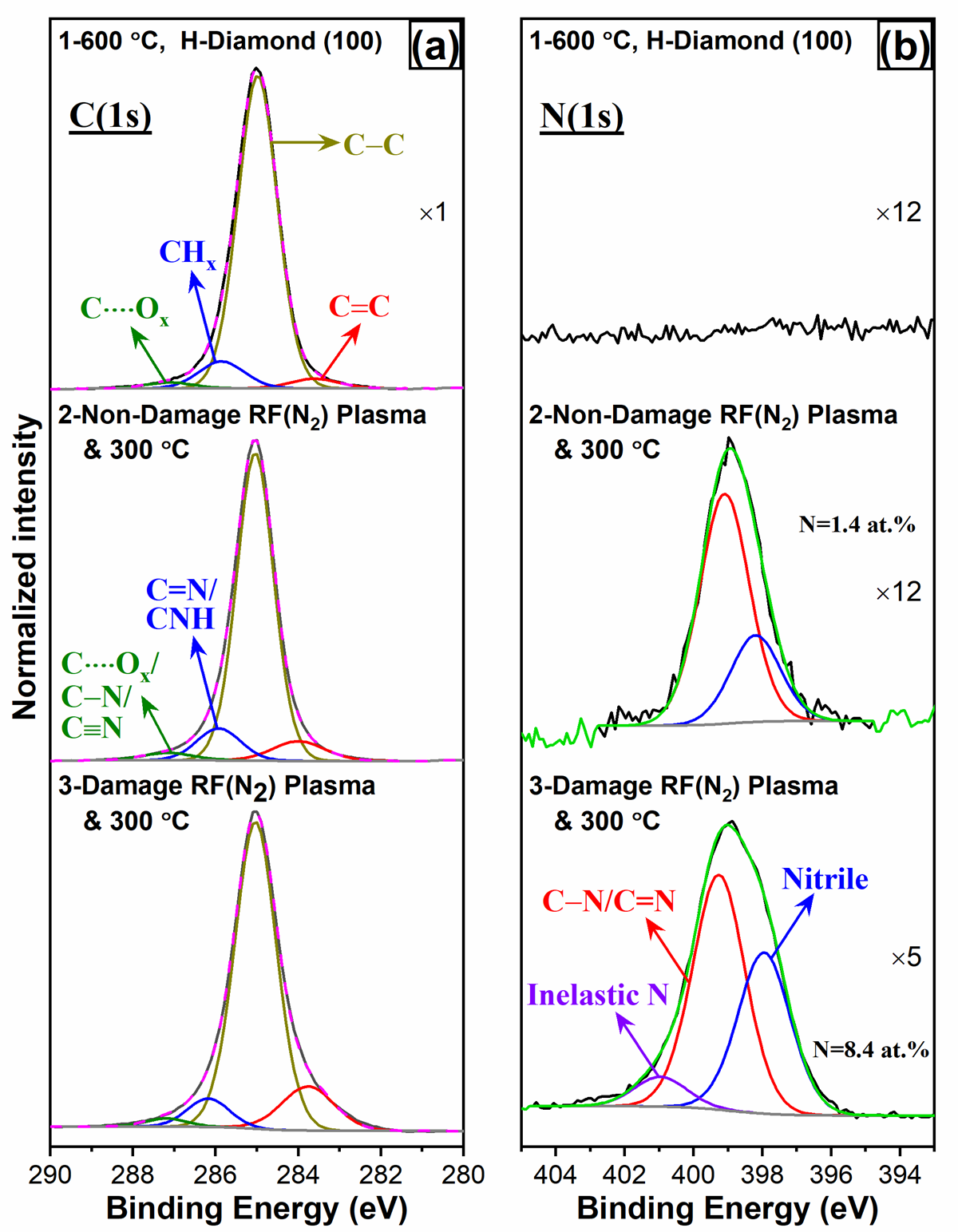}}
\caption{(a) C(1s) and (b) N(1s) XP spectra obtained from (1) 600 \textdegree C pristine; (2) non-damage RF(N$_{2}$) plasma exposed \& 300 \textdegree C annealed; and (3) damage RF(N$_{2}$) plasma exposed \& 300 \textdegree C annealed diamond (100) surfaces.}
\label{fig:XPspectra}
\end{figure}

Figure \ref{fig:XPspectra} (a) and (b) show the C(1s) and N(1s) XP spectra obtained from (1) 600 $^\circ$C annealed H-diamond; (2) non-damage RF(N$_{2}$) plasma exposed \& 300 \textdegree C annealed; and (3) damage RF(N$_{2}$) plasma exposed \& 300 \textdegree C annealed diamond (100) surfaces. In Figure 1(a), for 600 \textdegree C annealed H-diamond (100), the C(1s) XP line is located at 285.0 $\pm$ 0.1 eV (FWHM = 1.25 $\pm$ 0.05 eV) associated with sp$^{3}$ carbon bonds (C–C) in diamond \cite{RN1492, RN1544}. Deconvolution of C(1s) XP spectrum revealed three additional components positioned at 283.6 $\pm$ 0.2, 286.1 $\pm$ 0.2, eV and 287.2 $\pm$ 0.2 eV that are associated with sp$^{2}$ carbon bonds and non-diamond species (C=C) \cite{RN1484}, hydrogenated carbon on the surface (CH$_{x}$) \cite{RN1484}, and oxygen impurity on the surface carbon atom in C$\cdots$O$_{x}$ state \cite{RN1484}, which is due to partial oxidation of sample surface during the sample transfer into the UHV chamber, respectively. Upon the H-diamond (100) surface exposure to non-damage RF(N$_{2}$) plasma followed by vacuum annealing to 300 \textdegree C, the change in the C(1s) XP line shape is very minor. The C–C (sp$^{3}$) component slightly decreased while there was a minor increase in C=C (sp$^{2}$) component compared to the pristine H-diamond (100) surface. Spectral deconvolution shows minor increase in intensity of the 286.1 $\pm$ 0.2 eV and 287.2 $\pm$ 0.2 eV peaks. These two peaks have been previously attributed to carbon atoms bonded with nitrogen in two different bonding configurations: the 287.2 eV peak to C–N(ad)/C$\equiv$N (ad) bonds and some C$\cdots$O$_{x}$, and the peak at 286.1 eV to C=N(ad) along with CH$_{x}$ \cite{RN1370}. Upon the H-diamond (100) exposure to damage RF(N$_{2}$) plasma followed by vacuum annealing to 300 \textdegree C, the C(1s) XP line displays asymmetry in the lower binding energy side, indicating an increase in non-diamond content on the surface. Peak fitting reveals a substantial increase in C=C (sp$^{2}$) component compared to pristine H-diamond (100) surface while the CN associated components intensities are similar to non-damage RF(N$_{2}$) case. The increase in C=C (sp$^{2}$) concentration on the surface indicates severe damage induced to the diamond (100) surface upon exposure to damaging RF(N$_{2}$) plasma conditions. In Figure 1(b), the N(1s) XP line at 399.0 $\pm$ 0.2 eV is narrower for non-damage RF(N$_{2}$) plasma exposed diamond (100) surface compared with the one exposed to damage RF(N$_{2}$) plasma indicating differences in nitrogen bonding configuration and their relative concentrations. The estimated nitrogen concentration (at.\%), assuming homogeneous distribution within the bulk diamond, on non-damage and damage RF(N$_{2}$) plasma exposed diamond (100) surfaces followed by annealing 300 \textdegree C were 1.4 $\pm$ 0.2 and 8.4 $\pm$ 0.2 at.\%, respectively. Deconvolution of N(1s) XP lines reveals that the nitrogen bonded to carbon in nitrile configuration (398.0 $\pm$ 0.2 eV) \cite{RN1313, RN1310} is relatively high for damage RF(N$_{2}$) plasma exposed diamond (100) surface. Whereas, on the non-damage RF(N$_{2}$) plasma exposed diamond (100) surface, the C‒N/C=N configuration (399.0 $\pm$ 0.2 eV) \cite{RN1313, RN1310} prevails. A higher nitrile content in the former case is an indication of substantial damage induced to the diamond surface.

\begin{figure}[tbh]
{\includegraphics[width = 0.9 \linewidth]{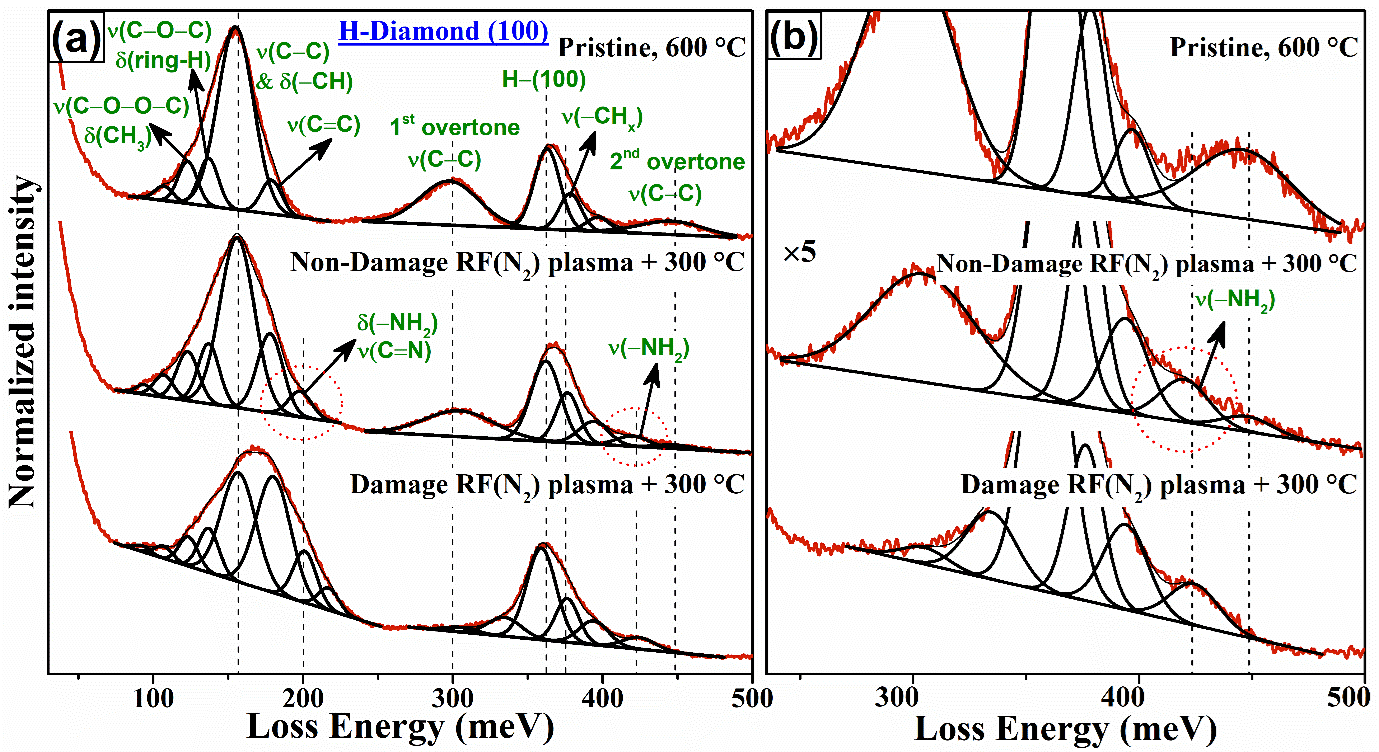}}
\caption{(a) HREEL spectra obtained from 600 \textdegree C annealed pristine, non-damage RF(N$_{2}$) plasma exposed \& 300 \textdegree C annealed and damage RF(N$_{2}$) plasma exposed \& 300 \textdegree C annealed diamond (100) surfaces. (b) ×5 magnification of the spectra presented in Figure \ref{fig:HREEL}(a) in the range of 240 to 500 meV.}
\label{fig:HREEL}
\end{figure}

HREEL spectroscopy is a high surface sensitive technique, which utilizes the inelastic scattering of electrons from a solid surface, to study the vibrational modes of surface or adsorbed species on a surface. Unlike other spectroscopic methods, HREEL spectroscopy is very sensitive to the upper atomic layer, in particular to the presence of hydrogen bonding. Figure \ref{fig:HREEL}(a) shows the HREEL spectra obtained from 600 \textdegree C annealed pristine, non-damage RF(N$_{2}$) plasma exposed \& 300 \textdegree C annealed and damage RF(N$_{2}$) plasma exposed \& 300 \textdegree C annealed diamond (100) surfaces. Figure \ref{fig:HREEL}(b) shows ×5 magnification of the spectra presented in Figure \ref{fig:HREEL}(a) in the range of 240 to 500 meV. 
In Figure \ref{fig:HREEL}(a), spectrum-1 depicts the typical HREEL spectrum obtained from H-diamond (100) surface shows two major peaks positioned at ~155 and ~360 meV, which are associated with mixed fundamental optical phonon of sp$^{3}$ bonded C‒C stretching vibrations and C‒H bending modes and C‒H stretching vibrations of sp$^{3}$ bonded C‒H species on the diamond surface, respectively. The other two broad and low-intensity peaks at ~303 and 450 meV are the 1$^{st}$ and 2$^{nd}$ overtones of optical phonons of sp$^{3}$ C‒C stretching vibrations of diamond carbons. These are signature peaks of a highly crystalline hydrogenated diamond \cite{RN1302, RN1273}, confirming that the measured diamond surface possesses high crystallinity. 

Deconvolution of the 155 meV peak in the HREEL spectrum of 600 \textdegree C annealed pristine H-diamond (100) reveals four peaks positioned at ~105, ~125, ~135, and ~180 eV associated with twisting and rocking deformation of C‒H bonds, ‒CH$_{3}$ rocking, HC=CH wagging vibrations, C‒H bending deformation and C=C stretching vibrations related to defects \cite{RN1315, RN1394}, respectively, in addition to 155 meV peak. Similarly, deconvolution of 360 meV peak reveals additional peaks at ~374 and 390 meV associated with stretching modes of sp$^{2}$ and sp hybridized CH species \cite{RN1302, RN1273}.

The HREEL spectra obtained from the non-damage and damage RF(N$_{2}$) plasma exposure followed by 300 \textdegree C annealed diamond (100) surfaces show few distinct spectral changes such as broadening of 155 meV peak, decrease/absence of 303 and 450 meV peaks associated with the 1$^{st}$ and 2$^{nd}$ order phonons and appearance of additional low intensity peak at ~420 meV. The new peak at ~420 meV (marked in red-dashed circle) is associated with ‒NH stretching mode. \cite{RN1361} To visualize this spectral feature, ×5 magnification of the spectra presented in Figure 2(a) in the range of 240 to 500 meV is shown in Figure \ref{fig:HREEL}(b). The absence of 303 and 450 meV peaks for damaging RF(N$_{2}$) plasma exposed surface confirms loss of structural order in the upper atomic layers of the diamond (100). Whereas, presence of 303 meV peak for non-damaging RF(N$_{2}$) plasma exposed surface corroborates retention of its structural order. Fitting of broad 155 meV peaks for non-damage and damage RF(N$_{2}$) plasma exposed surfaces reveals an increase in C=C content (compared to H-diamond) and presence of C=N bonding on the surface (indicated in a red-dashed circle).

From XPS and HREELS analyses, it can be concluded that H-diamond (100) surface exposure to non-damaging RF(N$_{2}$) plasma would result in nitrogen termination of diamond surface with minimal defects.

\subsection{Quantum properties of shallow NVs}
The NV centers themselves were characterized through optical fluorescence and quantum spin properties using a home-built confocal microscope.

\begin{figure}[htb]
{\includegraphics[width = 0.9 \linewidth]{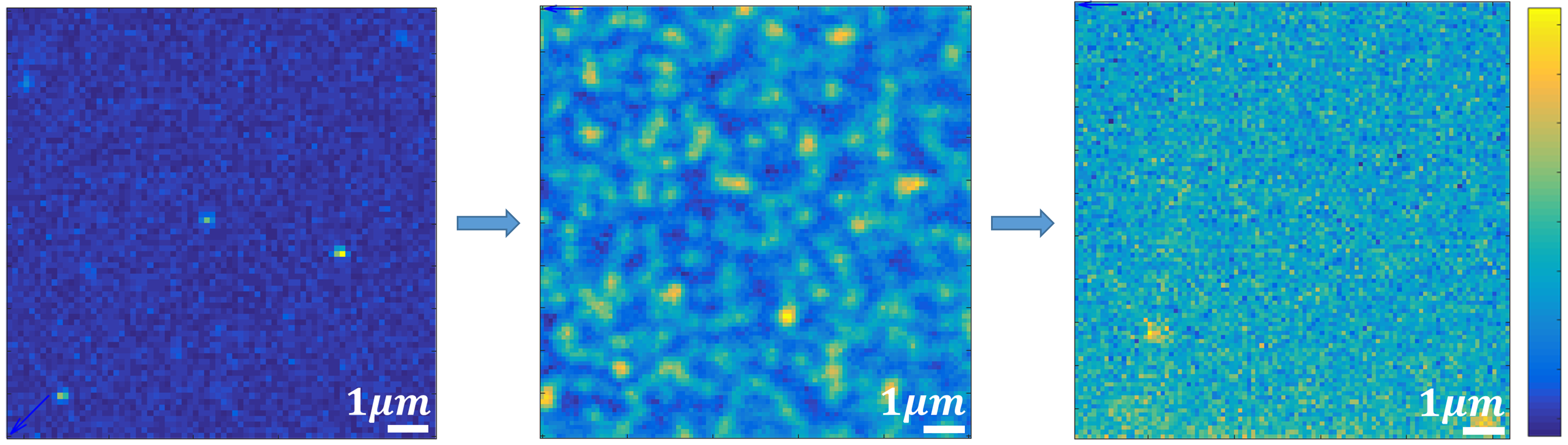}}
\caption{2D scans for the different surface treatments, each normalized by its own countrate (yellow – max. countrate, blue – min. countrate). Left to right: oxidation by acid, non-damaging $^{14}$N RF plasma, damaging $^{14}$N RF plasma. Both the damaging $^{14}$N RF plasma and the oxidation resulted in unstable NVs. Fluorescent spots seen in both cases did not generate any spin signal and disappeared in subsequent scans. After surface treatment using non-damaging $^{14}$N RF plasma conditions, NV stability increased along with the countrate and lifetime.}
\label{fig:FLscans}
\end{figure}

In Fig. \ref{fig:FLscans} we present optical fluorescence (FL) scans of the NV layer, comparing the untreated, acid cleaned surface, to surface termination with both damaging and non-damaging N plasma.
Both acid cleaning (oxidation) and damaging RF conditions resulted in unstable NVs, while the non-damaging RF conditions yielded a dense 2D scan of stable NVs. The following characterization of quantum spin properties were done under the non-damaging RF plasma conditions only, as under the other conditions the NVs were not stable enough to provide any signal.

\begin{figure}[tbh]
{\includegraphics[width = 0.9 \linewidth]{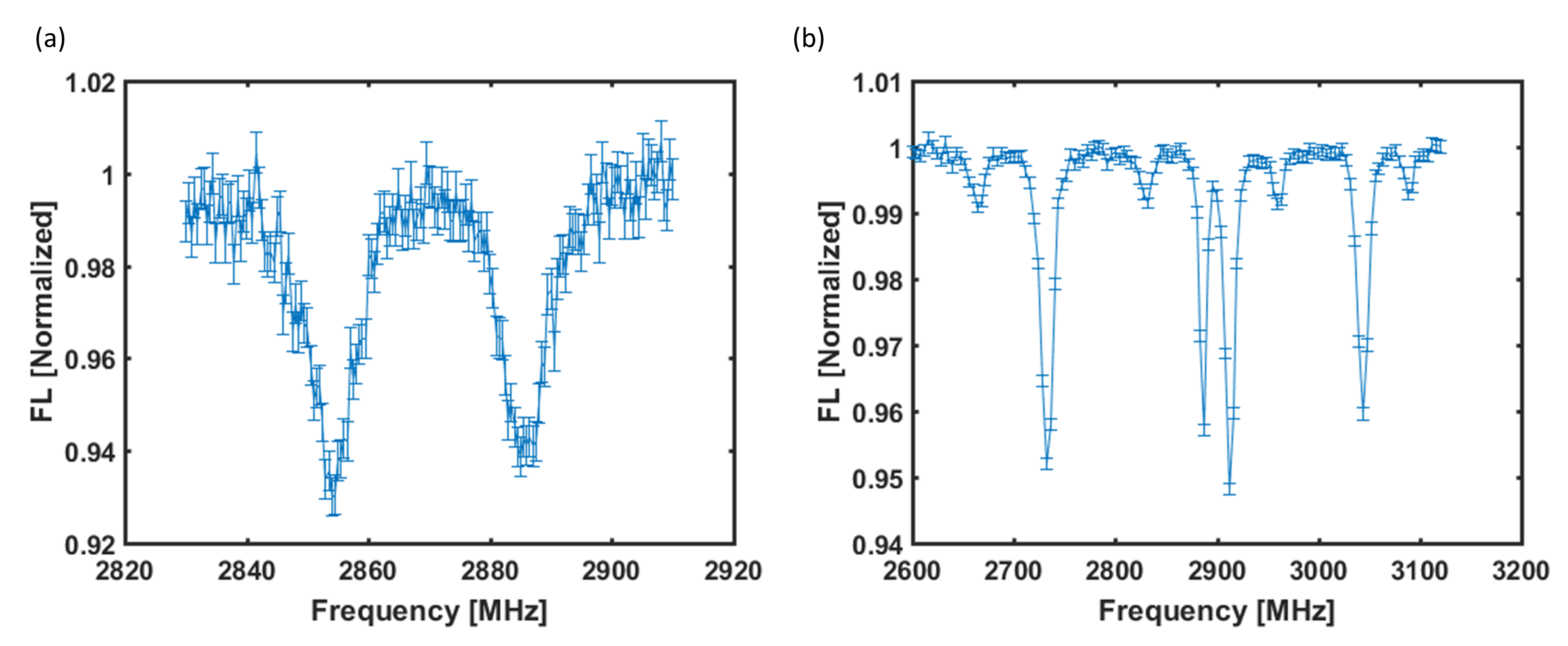}}
\caption{ESR measurements. (a) ESR measurement under low magnetic field oriented along the z axis (out of plane). (b) ESR measurement under a magnetic field such that the resonant frequencies of the different orientations split. We note that in certain cases multiple NV orientations are found, as shown in (b), although the implantation parameters were chosen to result in low NV concentration.}
\label{fig:ODMRs}
\end{figure}

Following the non-damaging RF plasma, the quantum spin properties of the shallow NVs were characterized by means of spin contrast and coherence measurements. We first measured the electron spin resonance (ESR) response, confirming the expected linear shift of the spin resonance frequencies as a function of the externally applied magnetic field (Zeeman shift), presented in Fig. \ref{fig:ODMRs}.

\begin{figure}[tbh]
{\includegraphics[width = 0.9 \linewidth]{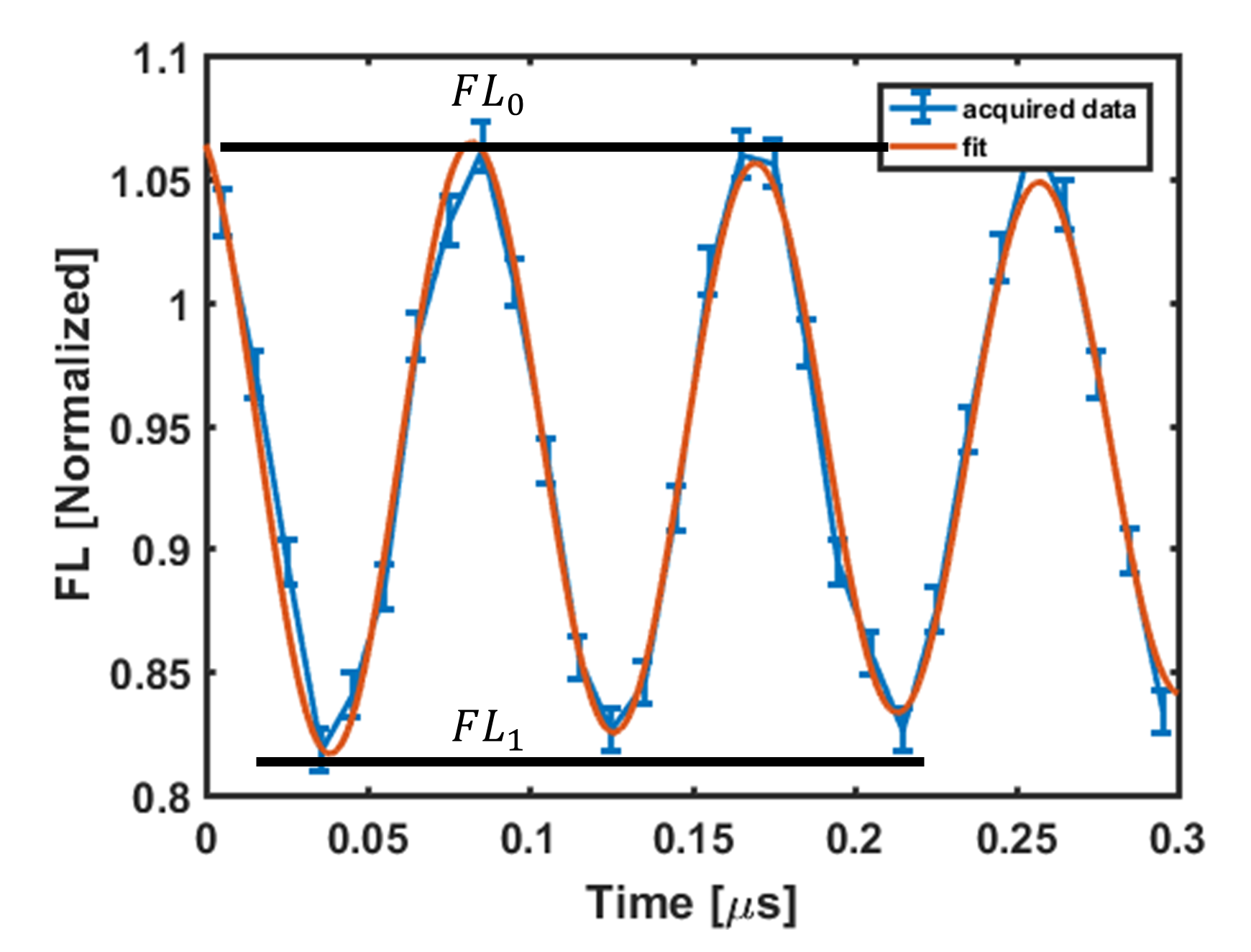}}
\caption{Typical Rabi driving measurement with an extracted contrast $C=0.26 \pm 0.005$. Maximal normalized fluorescence intensity corresponds to the 0 state and the minimal intensity to the 1 state.}
\label{fig:rabi}
\end{figure}

We define our spin contrast C as $C = FL_0 - FL_1$, where $FL_{0,1}$ are the normalized fluorescence intensities for the $0,1$ states of the NV spin. This corresponds to the high and low fluorescence values in a resonant driving experiment, i.e. Rabi driving, as presented in Fig. \ref{fig:rabi}. This contrast value corresponds to the NV charge stability \cite{kageura2017effect, yamano2017charge}, and we obtain following nitridation with non-damaging conditions a contrast of $C=0.21 \pm 0.006$.\\

\begin{figure}[tbh]
{\includegraphics[width = 0.9 \linewidth]{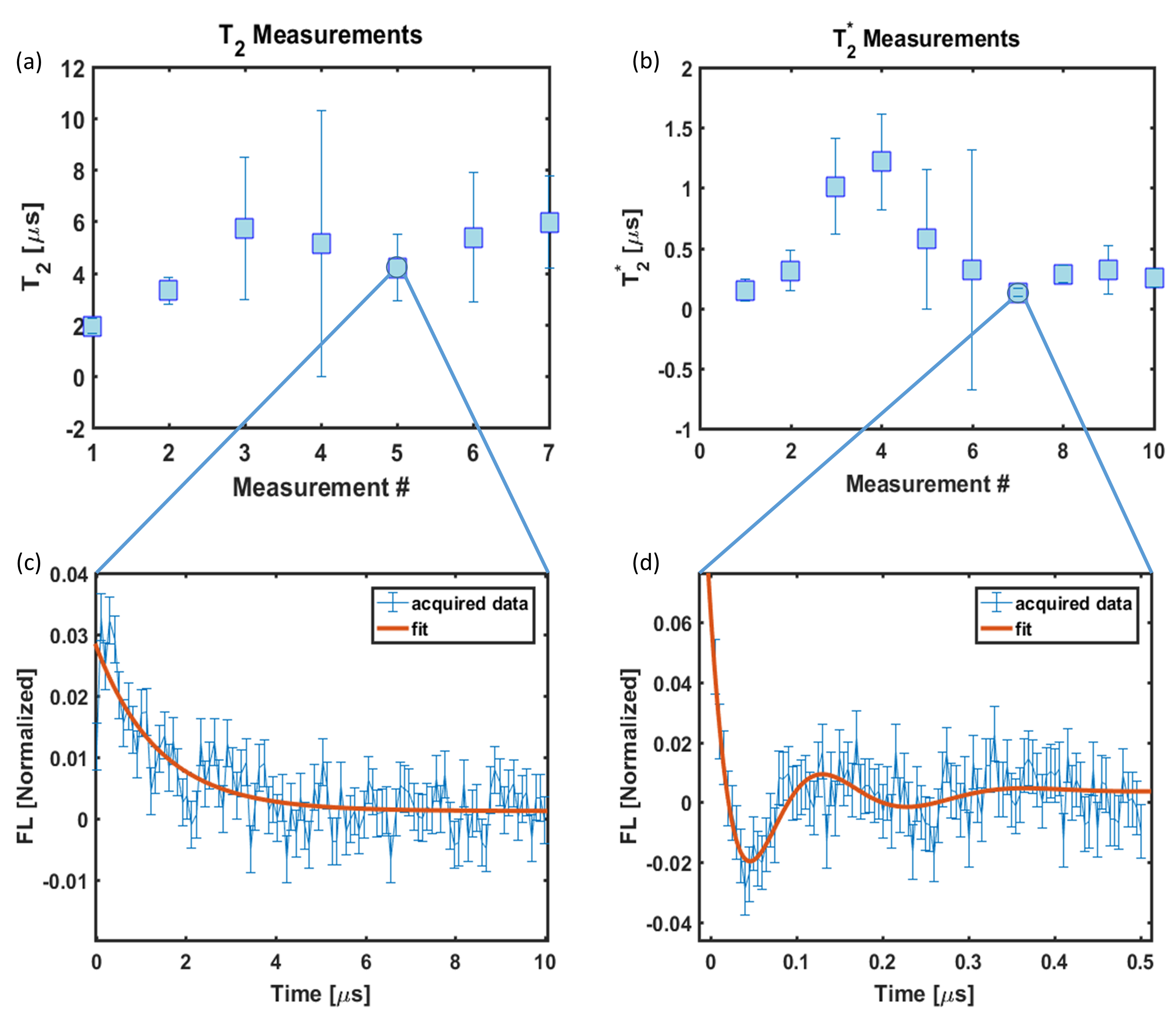}}
\caption{$T_2$ and $T_2^*$ coherence time values are summarized in (a) and (b) respectively. Characteristic measurements are shown in (c) and (d), with (c) presenting a $T_2$ measurement with a coherence time of $T_2 \simeq 4.2 \pm 0.65\ \mu$s and (d) presenting a $T_2^*$ measurement with coherence time of $T_2^* \simeq 0.13 \pm 0.02\ \mu$s. Full measurement results for the coherence times are summarized in table \ref{table:T2}.}
\label{fig:coherence}
\end{figure}

Coherence measurements were done using both Hahn-Echo and Ramsey FID sequences \cite{bargill2013coherence, childress2006coherent} to determine the $T_2$ and $T_2^*$ decay times respectively. The measured data (normalized fluorescence signal $S_{echo}$) is fitted for the Hahn-Echo measurements using
\begin{equation}
    S_{echo} = C e^{\left( - \frac{2 \tau}{T_2} \right)^3} + d,
\end{equation}
where C (contrast), d (background noise) and $T_2$ (coherence) are extracted from fits to the data. Ramsey measurements were performed off-resonance to enable more consistent fits to the decaying envelope, and were fitted to the normalized signal $S_{FID}$
\begin{eqnarray}
    S_{FID} &=& e^{\left( - \frac{\tau}{T_2^*} \right) ^2} \times \nonumber \\
    & \times &  \left \{ C_0 \cdot Cos(\pi f t) + C_1 \cdot Cos \left[ \pi (f+3) t \right] +\phi \right \}] + \nonumber \\
    &+& d,
\end{eqnarray}
where the contrasts $C_0$ and $C_1$, the phase $\phi$, the background noise $d$, the detuning $f$ and the coherence time $T_2^*$ are extracted from the fits to the data. The power of the stretched exponential decay of the envelope was chosen to be fixed (at $2$ for FID and at $3$ for echo) according to the expected theoretical dependence \cite{deSousa2009, stanwix2010exponent} and to reduce the number of fitting parameters.
The results obtained for several different NVs measured are summarized in table \ref{table:T2} and in Fig. \ref{fig:coherence}, depicting average coherence times of $T_2^* \simeq 0.46 \pm 0.24\, \mu$s and $T_2 \simeq 4.53 \pm 1.29\, \mu$s.

\begin{table}
\begin{tabularx}{0.9 \linewidth}
{ 
  | >{\raggedright\arraybackslash}X 
  || >{\centering\arraybackslash}X 
  | >{\centering\arraybackslash}X 
  | >{\raggedleft\arraybackslash}X |}
 \hline
 NV & $T_2 [\mu$s] & $T_2^* [\mu$s] & Magnetic Field [G] \\
 \hline
 \hline
 1 & 1.94 & 0.15 & 100 \\
 2 & 3.31 & 1.01  & 100 \\
 3 & 5.72 & 0.57 & 100 \\
 4 & 5.15 & 0.32  & 300 \\
 5 & 4.2 & 0.13 & 300 \\
 6 & 5.39 & 0.32 & 300 \\
 7 & 5.97 & 0.25 & 300\\
 \hline
\end{tabularx}

 \caption{Summary of Ramsey FID and Echo measurements on different FL spots.}
\label{table:T2}

\end{table}

\section{Summary and discussion}
In this work we addressed the issues arising in utilizing very shallow NVs ($\approx 3$ nm from the surface) for sensing applications. Such shallow NVs tend to experience significant noise from the surface (associated with impurities, dangling bonds and charge traps), resulting in limited charge/photo stability and reduced coherence times.

We characterized several shallow NVs created using 2 keV ion implantation, which demonstrated highly unstable fluorescence, essentially prohibiting spin measurements (ODMR, coherence). 

We then terminated the surface, first with an ultraclean plasma nitridation process in non-damaging conditions, followed by a similar process at higher energy, damaging conditions.   

Careful analysis of the surface using extensive in-situ spectroscopy, including XPS and HREELS, showed clean nitrogen termination of the surface using non-damaging conditions plasma, while exposing the surface to plasma in damaging conditions introduced substantial defects.

Finally, fluorescence spin measurements of the shallow NVs indicated little improvement in the damaging case, while for non-damaging conditions the fluorescence of the NVs was stabilized, enabling complete measurements of the spin properties (ODMR, Rabi driving and coherence). The NVs exhibited useful coherence times of $T_2^* \simeq 460$ ns and $T_2 \simeq 4.53\ \mu$s.

Our results suggest a novel method for the stabilization of NVs situated a few nanometers from the surface with decent coherence properties. Specifically, using nitrogen plasma in non-damaging conditions produce negligible defects on the surface and results in enhanced NV properties, compared to previously used oxygen termination, which creates additional defects and destabilization of near-surface NVs through etching effects, and nitrogen plasma with damaging conditions. These properties could significantly improve sensing applications based on shallow NVs. Moreover, we expect similar results using this technique with nanodiamonds, which would further expand the toolbox of NV sensing.

\appendix
\section{Methods}
For XPS and HREELS surface analyses, single crystal (100)-oriented diamonds (8×8×0.5 mm$^{3}$, Element Six Ltd.) were used. The samples were cleaned in an ultrasonic bath for 10 minutes using acetone and ethanol followed by 4 $\mu$m thick homoepitaxial buffer layer growth in microwave chemical vapor deposition (MWCVD) (2.5 GHz, power 6000 W) system using hydrogen (H$_2$) and methane (CH$_4$) as precursor gases with a flow rate and gas pressure of 250:10 standard cubic centimeters per minute (sccm) (H$_2$:CH$_4$) and 150 Torr, respectively. The substrate temperature during growth was 900 \textdegree C. The buffer layer deposition was carried out at the Israeli Center of Advanced Diamond Technologies (ICDAT Ltd.).

The surface RF(N$_{2}$) plasma treatment and analysis were carried out in two ultrahigh vacuum (UHV) systems maintained at a base pressure of $5 \times 10^{-10}$ Torr, as previously described. The UHV system-1 is equipped with RF(N$_{2}$) plasma processing facility and the UHV system-2 is equipped with XPS and HREELS facilities.

Following the buffer layer deposition, the H-diamond (100) sample was transferred ex situ (exposed to ambient conditions) into UHV system-1. The diamond sample was mounted onto a custom-designed 2-in-1 sample holder and heater (Boralectric resistive heater) that facilitates in situ annealing. Next, the sample was vacuum annealed to 600 \textdegree C for 5 min followed by RF(N$_{2}$) plasma exposure for 30 min at two different N$_{2}$ gas pressures: $3 \times 10^{-2}$ (damaging condition) and $7 \times 10^{-2}$ Torr (non-damaging condition), and at plasma power of 36 W. The RF processing setup consists of a non-line-of-sight plasma source described in detail previously \cite{RN1362}. After plasma exposure, the RF nitrided diamond (100) samples were transferred ex situ into UHV system-2 to carry out XPS and HREELS measurements to evaluate their surface chemical and structural properties. 

The XPS measurements were performed using a non-monochromatic Mg K$\alpha$ anode X-ray source (XR50, SPECS) at an incident angle of 55\textdegree\ from the surface normal and a hemispherical analyzer (PHOIBOS 100, SPECS). The incident X-ray photons energy line width was 0.68 eV, as per the source specifications provided by the manufacturer (SPECS, Germany). The XP spectra were measured at room temperature (RT) in the 200-600 eV binding energy range to determine the chemical state of the nitride surfaces. In addition, high-resolution measurements were carried out in the N(1s) and C(1s) narrow spectral ranges. The pass energy and scan step used in the XPS measurements were 15 eV and 0.1 eV, respectively. To minimize the surface charging effects, a thin (100 $\mu$m) Molybdenum plate mask with a circular opening (diameter = 8 mm) was placed in contact with the sample surface and firmly connected to the sample holder’s electrical ground. Deconvolution of XPS peaks was performed after background subtraction as described by Shirley \cite{RN1489} and using a mixed Gaussian (Y\%)-Lorentzian(X\%) peak shape, defined in CasaXPS software (version 2.3.15) as GL(X\%). The C(1s) and N(1s) peaks are suited to the default GL(30) peak shape \cite{RN1490, RN1491}. From these measurements, it was determined that no impurities were present on the H-diamond surface within the sensitivity of the XPS (~0.5 at.\%). HREELS measurements were carried out using primary electron energy of 8.4 eV. The FWHM of the elastically scattered electron peak was 16 meV. The spectra were recorded up to loss energies of 500 meV in the specular geometry at an incident angle of 55\textdegree\ from the surface normal. All spectra were recorded in situ at RT under UHV conditions after annealing to 300 \textdegree C. The deconvolution of the HREEL spectra was performed by means of XPSPEAK (version 4.1) software.

For optical studies, electronic grade single crystal diamond (100) (1.6×1.6×0.1 mm$^{3}$, Element Six, UK) was used. The diamond was boiled in a tri-acid (HNO$_{3}$+H$_{2}$SO$_{4}$+HClO$_{4}$) mixture for 30 min at 140 \textdegree C, followed by boiling in water for 30 min to remove non-diamond species from the diamond surface if any. Next, the diamond sample was placed onto a custom-designed 2-in-1 sample holder and heater with a molybdenum mask having a circular opening (diameter = 1.5 mm) and transferred into UHV system-1 to carry out RF(N$_{2}$) plasma treatment. Prior to plasma nitridation, the sample was vacuum annealed to 800 \textdegree C for 5 min to remove oxygen impurities. Then, the sample was exposed to non-damaging RF(N$_{2}$) plasma for 30 min as described above and the sample was taken out without any further processing to perform optical measurements. After these measurements, the same sample was boiled in tri-acid as described above, followed by vacuum annealing to 800 \textdegree C for 5 min and exposed to damaging RF(N$_{2}$) plasma for 30 min as described above.

\section{Unexpected fluorescence and ODMR effects}
\begin{figure}[tbh]
{\includegraphics[width = 0.9 \linewidth]{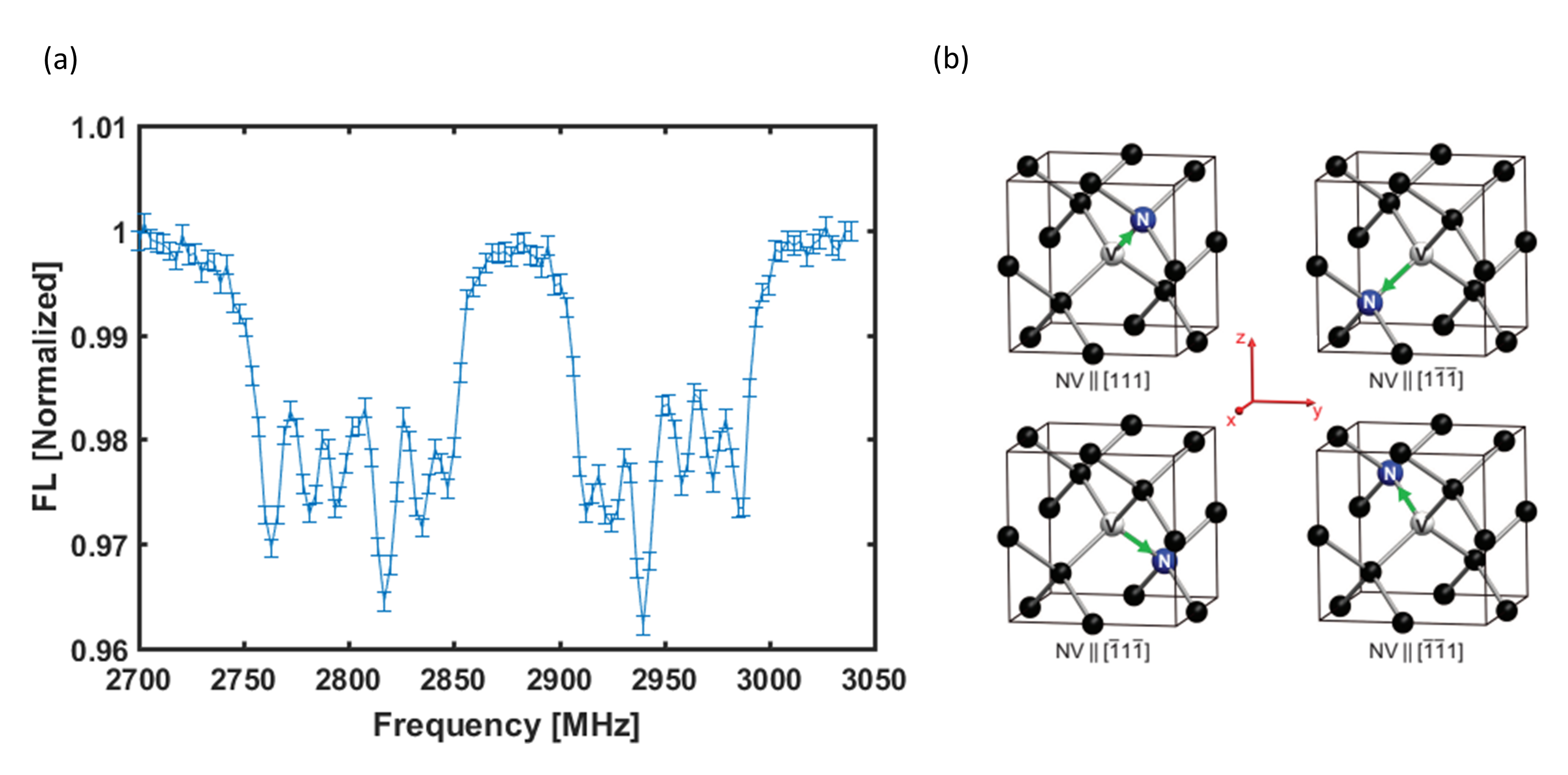}}
\caption{(a) ESR spectra taken from a fluorescent spot in a non-zero external magnetic field. (b) The possible orientations for a NV in a diamond (taken from \cite{pham2013thesis}). Each orientation contributes two resonant frequencies, associated with the $\pm \ket{1}$ spin states. As seen from (b) we expect to see up to 4 orientations (8 dips), but we see in (a) 12 dips.}
\label{fig:ESR}
\end{figure}

During the fluorescence measurement of the non-damaging termination we noticed unexpectedly high NV concentrations. The implantation parameters were chosen so as to create single NVs. However, the electron spin resonance (ESR) measurements revealed multiple resonant frequencies (Fig. \ref{fig:ESR}(a)). Given that each pair of resonance frequencies is attributed to a different NV orientation, we expect four possible pairs for a single crystal diamond sample according to the four possible orientations of NVs (Fig. \ref{fig:ESR}(b)). However, certain spots result in ESR scans that fit to 6 different orientations (Fig. \ref{fig:ESR}(a)). 

\begin{figure}[tbh]
{\includegraphics[width = 0.9 \linewidth]{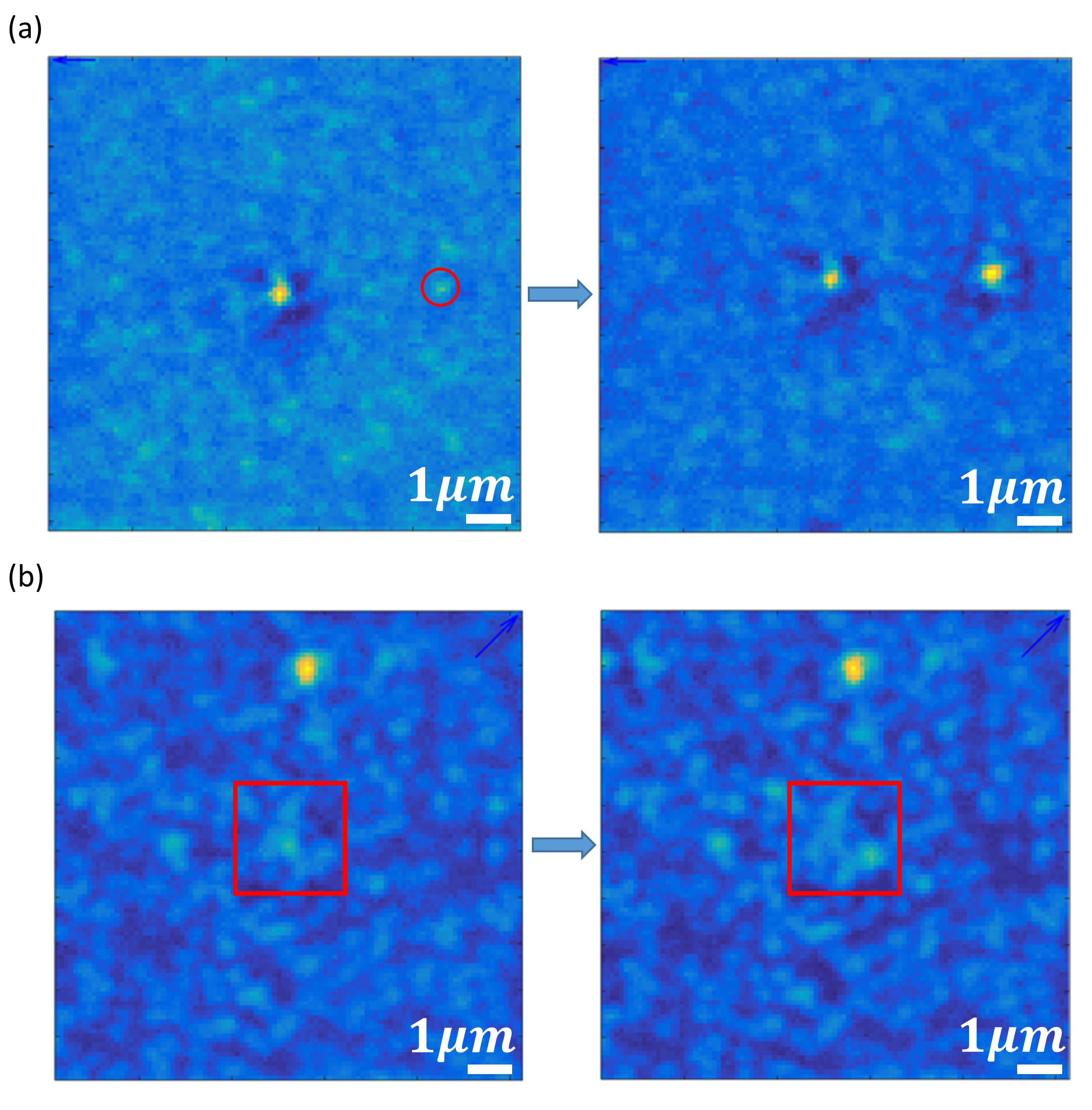}}
\caption{(a) FL spot fluorescence increases after running measurements on it. The red circle in the left scan depicts the FL spot which is going to be measured. The right scan shows how the fluorescence intensity changes at the FL spot and its surrounding area. (b) The left scan was taken before starting to measure the FL spot in the center of the square. The right scan shows how the fluorescence map changed after measurements. It can be seen that the center FL spot disappeared, while the surrounding area’s fluorescence intensity increased.}
\label{fig:FLchange}
\end{figure}

In addition, Fig. \ref{fig:FLchange} depicts a change in NV fluorescence observed due to repeated scans. We noticed variations in fluorescence intensity of NVs following different measurement sequences. Usually the NV emitted count rate would increase after running ESR and Rabi on the FL spot, while the FL in the surrounding area decreased. On fewer occasions we noticed the opposite phenomena – the central FL spot dimmed, while the FL of the surrounding area increased. This phenomenon persisted after annealing the diamond for 30 minutes (at both 300 \textdegree C and 400 \textdegree C).

\bibliography{ref.bib}

\begin{thebibliography}{36}%
\makeatletter
\providecommand \@ifxundefined [1]{%
 \@ifx{#1\undefined}
}%
\providecommand \@ifnum [1]{%
 \ifnum #1\expandafter \@firstoftwo
 \else \expandafter \@secondoftwo
 \fi
}%
\providecommand \@ifx [1]{%
 \ifx #1\expandafter \@firstoftwo
 \else \expandafter \@secondoftwo
 \fi
}%
\providecommand \natexlab [1]{#1}%
\providecommand \enquote  [1]{``#1''}%
\providecommand \bibnamefont  [1]{#1}%
\providecommand \bibfnamefont [1]{#1}%
\providecommand \citenamefont [1]{#1}%
\providecommand \href@noop [0]{\@secondoftwo}%
\providecommand \href [0]{\begingroup \@sanitize@url \@href}%
\providecommand \@href[1]{\@@startlink{#1}\@@href}%
\providecommand \@@href[1]{\endgroup#1\@@endlink}%
\providecommand \@sanitize@url [0]{\catcode `\\12\catcode `\$12\catcode
  `\&12\catcode `\#12\catcode `\^12\catcode `\_12\catcode `\%12\relax}%
\providecommand \@@startlink[1]{}%
\providecommand \@@endlink[0]{}%
\providecommand \url  [0]{\begingroup\@sanitize@url \@url }%
\providecommand \@url [1]{\endgroup\@href {#1}{\urlprefix }}%
\providecommand \urlprefix  [0]{URL }%
\providecommand \Eprint [0]{\href }%
\providecommand \doibase [0]{https://doi.org/}%
\providecommand \selectlanguage [0]{\@gobble}%
\providecommand \bibinfo  [0]{\@secondoftwo}%
\providecommand \bibfield  [0]{\@secondoftwo}%
\providecommand \translation [1]{[#1]}%
\providecommand \BibitemOpen [0]{}%
\providecommand \bibitemStop [0]{}%
\providecommand \bibitemNoStop [0]{.\EOS\space}%
\providecommand \EOS [0]{\spacefactor3000\relax}%
\providecommand \BibitemShut  [1]{\csname bibitem#1\endcsname}%
\let\auto@bib@innerbib\@empty
\bibitem [{\citenamefont {Degen}\ \emph {et~al.}(2017)\citenamefont {Degen},
  \citenamefont {Reinhard},\ and\ \citenamefont
  {Cappellaro}}]{degen2017review}%
  \BibitemOpen
  \bibfield  {author} {\bibinfo {author} {\bibfnamefont {C.~L.}\ \bibnamefont
  {Degen}}, \bibinfo {author} {\bibfnamefont {F.}~\bibnamefont {Reinhard}},\
  and\ \bibinfo {author} {\bibfnamefont {P.}~\bibnamefont {Cappellaro}},\
  }\bibfield  {title} {\bibinfo {title} {Quantum sensing},\ }\href
  {https://doi.org/10.1103/RevModPhys.89.035002} {\bibfield  {journal}
  {\bibinfo  {journal} {Reviews of modern physics}\ }\textbf {\bibinfo {volume}
  {89}},\ \bibinfo {pages} {035002} (\bibinfo {year} {2017})}\BibitemShut
  {NoStop}%
\bibitem [{\citenamefont {Taylor}\ \emph {et~al.}(2008)\citenamefont {Taylor},
  \citenamefont {Cappellaro}, \citenamefont {Childress}, \citenamefont {Jiang},
  \citenamefont {Budker}, \citenamefont {Hemmer}, \citenamefont {Yacoby},
  \citenamefont {Walsworth},\ and\ \citenamefont
  {Lukin}}]{taylor2008magnetometer}%
  \BibitemOpen
  \bibfield  {author} {\bibinfo {author} {\bibfnamefont {J.~M.}\ \bibnamefont
  {Taylor}}, \bibinfo {author} {\bibfnamefont {P.}~\bibnamefont {Cappellaro}},
  \bibinfo {author} {\bibfnamefont {L.}~\bibnamefont {Childress}}, \bibinfo
  {author} {\bibfnamefont {L.}~\bibnamefont {Jiang}}, \bibinfo {author}
  {\bibfnamefont {D.}~\bibnamefont {Budker}}, \bibinfo {author} {\bibfnamefont
  {P.}~\bibnamefont {Hemmer}}, \bibinfo {author} {\bibfnamefont
  {A.}~\bibnamefont {Yacoby}}, \bibinfo {author} {\bibfnamefont
  {R.}~\bibnamefont {Walsworth}},\ and\ \bibinfo {author} {\bibfnamefont
  {M.}~\bibnamefont {Lukin}},\ }\bibfield  {title} {\bibinfo {title}
  {High-sensitivity diamond magnetometer with nanoscale resolution},\ }\href
  {https://doi.org/10.1038/nphys1075} {\bibfield  {journal} {\bibinfo
  {journal} {Nature Physics}\ }\textbf {\bibinfo {volume} {4}},\ \bibinfo
  {pages} {810} (\bibinfo {year} {2008})}\BibitemShut {NoStop}%
\bibitem [{\citenamefont {Acosta}\ \emph {et~al.}(2009)\citenamefont {Acosta},
  \citenamefont {Bauch}, \citenamefont {Ledbetter}, \citenamefont {Santori},
  \citenamefont {Fu}, \citenamefont {Barclay}, \citenamefont {Beausoleil},
  \citenamefont {Linget}, \citenamefont {Roch}, \citenamefont {Treussart},
  \citenamefont {Chemerisov}, \citenamefont {Gawlik},\ and\ \citenamefont
  {Budker}}]{acosta2009diamonds}%
  \BibitemOpen
  \bibfield  {author} {\bibinfo {author} {\bibfnamefont {V.~M.}\ \bibnamefont
  {Acosta}}, \bibinfo {author} {\bibfnamefont {E.}~\bibnamefont {Bauch}},
  \bibinfo {author} {\bibfnamefont {M.~P.}\ \bibnamefont {Ledbetter}}, \bibinfo
  {author} {\bibfnamefont {C.}~\bibnamefont {Santori}}, \bibinfo {author}
  {\bibfnamefont {K.-M.~C.}\ \bibnamefont {Fu}}, \bibinfo {author}
  {\bibfnamefont {P.~E.}\ \bibnamefont {Barclay}}, \bibinfo {author}
  {\bibfnamefont {R.~G.}\ \bibnamefont {Beausoleil}}, \bibinfo {author}
  {\bibfnamefont {H.}~\bibnamefont {Linget}}, \bibinfo {author} {\bibfnamefont
  {J.~F.}\ \bibnamefont {Roch}}, \bibinfo {author} {\bibfnamefont
  {F.}~\bibnamefont {Treussart}}, \bibinfo {author} {\bibfnamefont
  {S.}~\bibnamefont {Chemerisov}}, \bibinfo {author} {\bibfnamefont
  {W.}~\bibnamefont {Gawlik}},\ and\ \bibinfo {author} {\bibfnamefont
  {D.}~\bibnamefont {Budker}},\ }\bibfield  {title} {\bibinfo {title} {Diamonds
  with a high density of nitrogen-vacancy centers for magnetometry
  applications},\ }\href {https://doi.org/10.1103/PhysRevB.80.115202}
  {\bibfield  {journal} {\bibinfo  {journal} {Physical Review B}\ }\textbf
  {\bibinfo {volume} {80}},\ \bibinfo {pages} {115202} (\bibinfo {year}
  {2009})}\BibitemShut {NoStop}%
\bibitem [{\citenamefont {Doherty}\ \emph {et~al.}(2013)\citenamefont
  {Doherty}, \citenamefont {Manson}, \citenamefont {Delaney}, \citenamefont
  {Jelezko}, \citenamefont {Wrachtrup},\ and\ \citenamefont
  {Hollenberg}}]{doherty2013nitrogen}%
  \BibitemOpen
  \bibfield  {author} {\bibinfo {author} {\bibfnamefont {M.~W.}\ \bibnamefont
  {Doherty}}, \bibinfo {author} {\bibfnamefont {N.~B.}\ \bibnamefont {Manson}},
  \bibinfo {author} {\bibfnamefont {P.}~\bibnamefont {Delaney}}, \bibinfo
  {author} {\bibfnamefont {F.}~\bibnamefont {Jelezko}}, \bibinfo {author}
  {\bibfnamefont {J.}~\bibnamefont {Wrachtrup}},\ and\ \bibinfo {author}
  {\bibfnamefont {L.~C.}\ \bibnamefont {Hollenberg}},\ }\bibfield  {title}
  {\bibinfo {title} {The nitrogen-vacancy colour centre in diamond},\ }\href
  {https://doi.org/https://doi.org/10.1016/j.physrep.2013.02.001} {\bibfield
  {journal} {\bibinfo  {journal} {Physics Reports}\ }\textbf {\bibinfo {volume}
  {528}},\ \bibinfo {pages} {1} (\bibinfo {year} {2013})}\BibitemShut {NoStop}%
\bibitem [{\citenamefont {Bar-Gill}\ \emph {et~al.}(2013)\citenamefont
  {Bar-Gill}, \citenamefont {Pham}, \citenamefont {Jarmola}, \citenamefont
  {Budker},\ and\ \citenamefont {Walsworth}}]{bargill2013coherence}%
  \BibitemOpen
  \bibfield  {author} {\bibinfo {author} {\bibfnamefont {N.}~\bibnamefont
  {Bar-Gill}}, \bibinfo {author} {\bibfnamefont {L.~M.}\ \bibnamefont {Pham}},
  \bibinfo {author} {\bibfnamefont {A.}~\bibnamefont {Jarmola}}, \bibinfo
  {author} {\bibfnamefont {D.}~\bibnamefont {Budker}},\ and\ \bibinfo {author}
  {\bibfnamefont {R.~L.}\ \bibnamefont {Walsworth}},\ }\bibfield  {title}
  {\bibinfo {title} {Solid-state electronic spin coherence time approaching one
  second},\ }\href {https://doi.org/10.1038/ncomms2771} {\bibfield  {journal}
  {\bibinfo  {journal} {Nature communications}\ }\textbf {\bibinfo {volume}
  {4}},\ \bibinfo {pages} {1} (\bibinfo {year} {2013})}\BibitemShut {NoStop}%
\bibitem [{\citenamefont {Childress}\ \emph {et~al.}(2006)\citenamefont
  {Childress}, \citenamefont {Gurudev~Dutt}, \citenamefont {Taylor},
  \citenamefont {Zibrov}, \citenamefont {Jelezko}, \citenamefont {Wrachtrup},
  \citenamefont {Hemmer},\ and\ \citenamefont {Lukin}}]{childress2006coherent}%
  \BibitemOpen
  \bibfield  {author} {\bibinfo {author} {\bibfnamefont {L.}~\bibnamefont
  {Childress}}, \bibinfo {author} {\bibfnamefont {M.}~\bibnamefont
  {Gurudev~Dutt}}, \bibinfo {author} {\bibfnamefont {J.}~\bibnamefont
  {Taylor}}, \bibinfo {author} {\bibfnamefont {A.}~\bibnamefont {Zibrov}},
  \bibinfo {author} {\bibfnamefont {F.}~\bibnamefont {Jelezko}}, \bibinfo
  {author} {\bibfnamefont {J.}~\bibnamefont {Wrachtrup}}, \bibinfo {author}
  {\bibfnamefont {P.}~\bibnamefont {Hemmer}},\ and\ \bibinfo {author}
  {\bibfnamefont {M.}~\bibnamefont {Lukin}},\ }\bibfield  {title} {\bibinfo
  {title} {Coherent dynamics of coupled electron and nuclear spin qubits in
  diamond},\ }\href {https://doi.org/10.1126/science.1131871} {\bibfield
  {journal} {\bibinfo  {journal} {Science}\ }\textbf {\bibinfo {volume}
  {314}},\ \bibinfo {pages} {281} (\bibinfo {year} {2006})}\BibitemShut
  {NoStop}%
\bibitem [{\citenamefont {Romach}\ \emph {et~al.}(2015)\citenamefont {Romach},
  \citenamefont {M{\"u}ller}, \citenamefont {Unden}, \citenamefont {Rogers},
  \citenamefont {Isoda}, \citenamefont {Itoh}, \citenamefont {Markham},
  \citenamefont {Stacey}, \citenamefont {Meijer}, \citenamefont {Pezzagna}
  \emph {et~al.}}]{romach2015spectroscopy}%
  \BibitemOpen
  \bibfield  {author} {\bibinfo {author} {\bibfnamefont {Y.}~\bibnamefont
  {Romach}}, \bibinfo {author} {\bibfnamefont {C.}~\bibnamefont {M{\"u}ller}},
  \bibinfo {author} {\bibfnamefont {T.}~\bibnamefont {Unden}}, \bibinfo
  {author} {\bibfnamefont {L.}~\bibnamefont {Rogers}}, \bibinfo {author}
  {\bibfnamefont {T.}~\bibnamefont {Isoda}}, \bibinfo {author} {\bibfnamefont
  {K.}~\bibnamefont {Itoh}}, \bibinfo {author} {\bibfnamefont {M.}~\bibnamefont
  {Markham}}, \bibinfo {author} {\bibfnamefont {A.}~\bibnamefont {Stacey}},
  \bibinfo {author} {\bibfnamefont {J.}~\bibnamefont {Meijer}}, \bibinfo
  {author} {\bibfnamefont {S.}~\bibnamefont {Pezzagna}}, \emph {et~al.},\
  }\bibfield  {title} {\bibinfo {title} {Spectroscopy of surface-induced noise
  using shallow spins in diamond},\ }\href
  {https://doi.org/10.1103/PhysRevLett.114.017601} {\bibfield  {journal}
  {\bibinfo  {journal} {Physical review letters}\ }\textbf {\bibinfo {volume}
  {114}},\ \bibinfo {pages} {017601} (\bibinfo {year} {2015})}\BibitemShut
  {NoStop}%
\bibitem [{\citenamefont {Stacey}\ \emph {et~al.}(2019)\citenamefont {Stacey},
  \citenamefont {Dontschuk}, \citenamefont {Chou}, \citenamefont {Broadway},
  \citenamefont {Schenk}, \citenamefont {Sear}, \citenamefont {Tetienne},
  \citenamefont {Hoffman}, \citenamefont {Prawer}, \citenamefont {Pakes} \emph
  {et~al.}}]{stacey2019sp2}%
  \BibitemOpen
  \bibfield  {author} {\bibinfo {author} {\bibfnamefont {A.}~\bibnamefont
  {Stacey}}, \bibinfo {author} {\bibfnamefont {N.}~\bibnamefont {Dontschuk}},
  \bibinfo {author} {\bibfnamefont {J.-P.}\ \bibnamefont {Chou}}, \bibinfo
  {author} {\bibfnamefont {D.~A.}\ \bibnamefont {Broadway}}, \bibinfo {author}
  {\bibfnamefont {A.~K.}\ \bibnamefont {Schenk}}, \bibinfo {author}
  {\bibfnamefont {M.~J.}\ \bibnamefont {Sear}}, \bibinfo {author}
  {\bibfnamefont {J.-P.}\ \bibnamefont {Tetienne}}, \bibinfo {author}
  {\bibfnamefont {A.}~\bibnamefont {Hoffman}}, \bibinfo {author} {\bibfnamefont
  {S.}~\bibnamefont {Prawer}}, \bibinfo {author} {\bibfnamefont {C.~I.}\
  \bibnamefont {Pakes}}, \emph {et~al.},\ }\bibfield  {title} {\bibinfo {title}
  {Evidence for primal sp2 defects at the diamond surface: candidates for
  electron trapping and noise sources},\ }\href
  {https://doi.org/https://doi.org/10.1002/admi.201801449} {\bibfield
  {journal} {\bibinfo  {journal} {Advanced Materials Interfaces}\ }\textbf
  {\bibinfo {volume} {6}},\ \bibinfo {pages} {1801449} (\bibinfo {year}
  {2019})}\BibitemShut {NoStop}%
\bibitem [{\citenamefont {Myers}\ \emph {et~al.}(2017)\citenamefont {Myers},
  \citenamefont {Ariyaratne},\ and\ \citenamefont
  {Jayich}}]{myers2017surfaceT1}%
  \BibitemOpen
  \bibfield  {author} {\bibinfo {author} {\bibfnamefont {B.~A.}\ \bibnamefont
  {Myers}}, \bibinfo {author} {\bibfnamefont {A.}~\bibnamefont {Ariyaratne}},\
  and\ \bibinfo {author} {\bibfnamefont {A.~B.}\ \bibnamefont {Jayich}},\
  }\bibfield  {title} {\bibinfo {title} {Double-quantum spin-relaxation limits
  to coherence of near-surface nitrogen-vacancy centers},\ }\href
  {https://doi.org/10.1103/PhysRevLett.118.197201} {\bibfield  {journal}
  {\bibinfo  {journal} {Physical review letters}\ }\textbf {\bibinfo {volume}
  {118}},\ \bibinfo {pages} {197201} (\bibinfo {year} {2017})}\BibitemShut
  {NoStop}%
\bibitem [{\citenamefont {Kim}\ \emph {et~al.}(2015)\citenamefont {Kim},
  \citenamefont {Mamin}, \citenamefont {Sherwood}, \citenamefont {Ohno},
  \citenamefont {Awschalom},\ and\ \citenamefont {Rugar}}]{kim2015decoherence}%
  \BibitemOpen
  \bibfield  {author} {\bibinfo {author} {\bibfnamefont {M.}~\bibnamefont
  {Kim}}, \bibinfo {author} {\bibfnamefont {H.}~\bibnamefont {Mamin}}, \bibinfo
  {author} {\bibfnamefont {M.}~\bibnamefont {Sherwood}}, \bibinfo {author}
  {\bibfnamefont {K.}~\bibnamefont {Ohno}}, \bibinfo {author} {\bibfnamefont
  {D.}~\bibnamefont {Awschalom}},\ and\ \bibinfo {author} {\bibfnamefont
  {D.}~\bibnamefont {Rugar}},\ }\bibfield  {title} {\bibinfo {title}
  {Decoherence of near-surface nitrogen-vacancy centers due to electric field
  noise},\ }\href {https://doi.org/10.1103/PhysRevLett.115.087602} {\bibfield
  {journal} {\bibinfo  {journal} {Physical review letters}\ }\textbf {\bibinfo
  {volume} {115}},\ \bibinfo {pages} {087602} (\bibinfo {year}
  {2015})}\BibitemShut {NoStop}%
\bibitem [{\citenamefont {Ofori-Okai}\ \emph {et~al.}(2012)\citenamefont
  {Ofori-Okai}, \citenamefont {Pezzagna}, \citenamefont {Chang}, \citenamefont
  {Loretz}, \citenamefont {Schirhagl}, \citenamefont {Tao}, \citenamefont
  {Moores}, \citenamefont {Groot-Berning}, \citenamefont {Meijer},\ and\
  \citenamefont {Degen}}]{ofori2012spin}%
  \BibitemOpen
  \bibfield  {author} {\bibinfo {author} {\bibfnamefont {B.}~\bibnamefont
  {Ofori-Okai}}, \bibinfo {author} {\bibfnamefont {S.}~\bibnamefont
  {Pezzagna}}, \bibinfo {author} {\bibfnamefont {K.}~\bibnamefont {Chang}},
  \bibinfo {author} {\bibfnamefont {M.}~\bibnamefont {Loretz}}, \bibinfo
  {author} {\bibfnamefont {R.}~\bibnamefont {Schirhagl}}, \bibinfo {author}
  {\bibfnamefont {Y.}~\bibnamefont {Tao}}, \bibinfo {author} {\bibfnamefont
  {B.}~\bibnamefont {Moores}}, \bibinfo {author} {\bibfnamefont
  {K.}~\bibnamefont {Groot-Berning}}, \bibinfo {author} {\bibfnamefont
  {J.}~\bibnamefont {Meijer}},\ and\ \bibinfo {author} {\bibfnamefont
  {C.}~\bibnamefont {Degen}},\ }\bibfield  {title} {\bibinfo {title} {Spin
  properties of very shallow nitrogen vacancy defects in diamond},\ }\href
  {https://doi.org/10.1103/PhysRevB.86.081406} {\bibfield  {journal} {\bibinfo
  {journal} {Physical Review B}\ }\textbf {\bibinfo {volume} {86}},\ \bibinfo
  {pages} {081406} (\bibinfo {year} {2012})}\BibitemShut {NoStop}%
\bibitem [{\citenamefont {Chou}\ \emph {et~al.}(2017)\citenamefont {Chou},
  \citenamefont {Retzker},\ and\ \citenamefont {Gali}}]{chou2017nitrogen}%
  \BibitemOpen
  \bibfield  {author} {\bibinfo {author} {\bibfnamefont {J.-P.}\ \bibnamefont
  {Chou}}, \bibinfo {author} {\bibfnamefont {A.}~\bibnamefont {Retzker}},\ and\
  \bibinfo {author} {\bibfnamefont {A.}~\bibnamefont {Gali}},\ }\bibfield
  {title} {\bibinfo {title} {Nitrogen-terminated diamond (111) surface for
  room-temperature quantum sensing and simulation},\ }\href
  {https://doi.org/10.1021/acs.nanolett.6b05023} {\bibfield  {journal}
  {\bibinfo  {journal} {Nano letters}\ }\textbf {\bibinfo {volume} {17}},\
  \bibinfo {pages} {2294} (\bibinfo {year} {2017})}\BibitemShut {NoStop}%
\bibitem [{\citenamefont {Kawai}\ \emph {et~al.}(2019)\citenamefont {Kawai},
  \citenamefont {Yamano}, \citenamefont {Sonoda}, \citenamefont {Kato},
  \citenamefont {Buendia}, \citenamefont {Kageura}, \citenamefont {Fukuda},
  \citenamefont {Okada}, \citenamefont {Tanii}, \citenamefont {Higuchi} \emph
  {et~al.}}]{kawai2019nitrogen}%
  \BibitemOpen
  \bibfield  {author} {\bibinfo {author} {\bibfnamefont {S.}~\bibnamefont
  {Kawai}}, \bibinfo {author} {\bibfnamefont {H.}~\bibnamefont {Yamano}},
  \bibinfo {author} {\bibfnamefont {T.}~\bibnamefont {Sonoda}}, \bibinfo
  {author} {\bibfnamefont {K.}~\bibnamefont {Kato}}, \bibinfo {author}
  {\bibfnamefont {J.~J.}\ \bibnamefont {Buendia}}, \bibinfo {author}
  {\bibfnamefont {T.}~\bibnamefont {Kageura}}, \bibinfo {author} {\bibfnamefont
  {R.}~\bibnamefont {Fukuda}}, \bibinfo {author} {\bibfnamefont
  {T.}~\bibnamefont {Okada}}, \bibinfo {author} {\bibfnamefont
  {T.}~\bibnamefont {Tanii}}, \bibinfo {author} {\bibfnamefont
  {T.}~\bibnamefont {Higuchi}}, \emph {et~al.},\ }\bibfield  {title} {\bibinfo
  {title} {Nitrogen-terminated diamond surface for nanoscale nmr by shallow
  nitrogen-vacancy centers},\ }\href {https://doi.org/10.1021/acs.jpcc.8b11274}
  {\bibfield  {journal} {\bibinfo  {journal} {The Journal of Physical Chemistry
  C}\ }\textbf {\bibinfo {volume} {123}},\ \bibinfo {pages} {3594} (\bibinfo
  {year} {2019})}\BibitemShut {NoStop}%
\bibitem [{\citenamefont {Bluvstein}\ \emph
  {et~al.}(2019{\natexlab{a}})\citenamefont {Bluvstein}, \citenamefont
  {Zhang},\ and\ \citenamefont {Jayich}}]{bluvstein2019surfaceCharge}%
  \BibitemOpen
  \bibfield  {author} {\bibinfo {author} {\bibfnamefont {D.}~\bibnamefont
  {Bluvstein}}, \bibinfo {author} {\bibfnamefont {Z.}~\bibnamefont {Zhang}},\
  and\ \bibinfo {author} {\bibfnamefont {A.~C.~B.}\ \bibnamefont {Jayich}},\
  }\bibfield  {title} {\bibinfo {title} {Identifying and mitigating charge
  instabilities in shallow diamond nitrogen-vacancy centers},\ }\href
  {https://doi.org/10.1103/PhysRevLett.122.076101} {\bibfield  {journal}
  {\bibinfo  {journal} {Physical review letters}\ }\textbf {\bibinfo {volume}
  {122}},\ \bibinfo {pages} {076101} (\bibinfo {year}
  {2019}{\natexlab{a}})}\BibitemShut {NoStop}%
\bibitem [{\citenamefont {Bluvstein}\ \emph
  {et~al.}(2019{\natexlab{b}})\citenamefont {Bluvstein}, \citenamefont {Zhang},
  \citenamefont {McLellan}, \citenamefont {Williams},\ and\ \citenamefont
  {Jayich}}]{bluvstein2019extendingSurface}%
  \BibitemOpen
  \bibfield  {author} {\bibinfo {author} {\bibfnamefont {D.}~\bibnamefont
  {Bluvstein}}, \bibinfo {author} {\bibfnamefont {Z.}~\bibnamefont {Zhang}},
  \bibinfo {author} {\bibfnamefont {C.~A.}\ \bibnamefont {McLellan}}, \bibinfo
  {author} {\bibfnamefont {N.~R.}\ \bibnamefont {Williams}},\ and\ \bibinfo
  {author} {\bibfnamefont {A.~C.~B.}\ \bibnamefont {Jayich}},\ }\bibfield
  {title} {\bibinfo {title} {Extending the quantum coherence of a near-surface
  qubit by coherently driving the paramagnetic surface environment},\ }\href
  {https://doi.org/10.1103/PhysRevLett.123.146804} {\bibfield  {journal}
  {\bibinfo  {journal} {Physical review letters}\ }\textbf {\bibinfo {volume}
  {123}},\ \bibinfo {pages} {146804} (\bibinfo {year}
  {2019}{\natexlab{b}})}\BibitemShut {NoStop}%
\bibitem [{\citenamefont {Sangtawesin}\ \emph {et~al.}(2019)\citenamefont
  {Sangtawesin}, \citenamefont {Dwyer}, \citenamefont {Srinivasan},
  \citenamefont {Allred}, \citenamefont {Rodgers}, \citenamefont {De~Greve},
  \citenamefont {Stacey}, \citenamefont {Dontschuk}, \citenamefont {O'Donnell},
  \citenamefont {Hu}, \citenamefont {Evans}, \citenamefont {Jaye},
  \citenamefont {Fischer}, \citenamefont {Markham}, \citenamefont {Twitchen},
  \citenamefont {Park}, \citenamefont {Lukin},\ and\ \citenamefont
  {de~Leon}}]{sangtawesin2019origins}%
  \BibitemOpen
  \bibfield  {author} {\bibinfo {author} {\bibfnamefont {S.}~\bibnamefont
  {Sangtawesin}}, \bibinfo {author} {\bibfnamefont {B.~L.}\ \bibnamefont
  {Dwyer}}, \bibinfo {author} {\bibfnamefont {S.}~\bibnamefont {Srinivasan}},
  \bibinfo {author} {\bibfnamefont {J.~J.}\ \bibnamefont {Allred}}, \bibinfo
  {author} {\bibfnamefont {L.~V.~H.}\ \bibnamefont {Rodgers}}, \bibinfo
  {author} {\bibfnamefont {K.}~\bibnamefont {De~Greve}}, \bibinfo {author}
  {\bibfnamefont {A.}~\bibnamefont {Stacey}}, \bibinfo {author} {\bibfnamefont
  {N.}~\bibnamefont {Dontschuk}}, \bibinfo {author} {\bibfnamefont {K.~M.}\
  \bibnamefont {O'Donnell}}, \bibinfo {author} {\bibfnamefont {D.}~\bibnamefont
  {Hu}}, \bibinfo {author} {\bibfnamefont {D.~A.}\ \bibnamefont {Evans}},
  \bibinfo {author} {\bibfnamefont {C.}~\bibnamefont {Jaye}}, \bibinfo {author}
  {\bibfnamefont {D.~A.}\ \bibnamefont {Fischer}}, \bibinfo {author}
  {\bibfnamefont {M.~L.}\ \bibnamefont {Markham}}, \bibinfo {author}
  {\bibfnamefont {D.~J.}\ \bibnamefont {Twitchen}}, \bibinfo {author}
  {\bibfnamefont {H.}~\bibnamefont {Park}}, \bibinfo {author} {\bibfnamefont
  {M.~D.}\ \bibnamefont {Lukin}},\ and\ \bibinfo {author} {\bibfnamefont
  {N.~P.}\ \bibnamefont {de~Leon}},\ }\bibfield  {title} {\bibinfo {title}
  {Origins of diamond surface noise probed by correlating single-spin
  measurements with surface spectroscopy},\ }\href
  {https://doi.org/10.1103/PhysRevX.9.031052} {\bibfield  {journal} {\bibinfo
  {journal} {Physical Review X}\ }\textbf {\bibinfo {volume} {9}},\ \bibinfo
  {pages} {031052} (\bibinfo {year} {2019})}\BibitemShut {NoStop}%
\bibitem [{\citenamefont {Kageura}\ \emph {et~al.}(2017)\citenamefont
  {Kageura}, \citenamefont {Kato}, \citenamefont {Yamano}, \citenamefont
  {Suaebah}, \citenamefont {Kajiya}, \citenamefont {Kawai}, \citenamefont
  {Inaba}, \citenamefont {Tanii}, \citenamefont {Haruyama}, \citenamefont
  {Yamada} \emph {et~al.}}]{kageura2017effect}%
  \BibitemOpen
  \bibfield  {author} {\bibinfo {author} {\bibfnamefont {T.}~\bibnamefont
  {Kageura}}, \bibinfo {author} {\bibfnamefont {K.}~\bibnamefont {Kato}},
  \bibinfo {author} {\bibfnamefont {H.}~\bibnamefont {Yamano}}, \bibinfo
  {author} {\bibfnamefont {E.}~\bibnamefont {Suaebah}}, \bibinfo {author}
  {\bibfnamefont {M.}~\bibnamefont {Kajiya}}, \bibinfo {author} {\bibfnamefont
  {S.}~\bibnamefont {Kawai}}, \bibinfo {author} {\bibfnamefont
  {M.}~\bibnamefont {Inaba}}, \bibinfo {author} {\bibfnamefont
  {T.}~\bibnamefont {Tanii}}, \bibinfo {author} {\bibfnamefont
  {M.}~\bibnamefont {Haruyama}}, \bibinfo {author} {\bibfnamefont
  {K.}~\bibnamefont {Yamada}}, \emph {et~al.},\ }\bibfield  {title} {\bibinfo
  {title} {Effect of a radical exposure nitridation surface on the charge
  stability of shallow nitrogen-vacancy centers in diamond},\ }\href
  {https://doi.org/10.7567/APEX.10.055503} {\bibfield  {journal} {\bibinfo
  {journal} {Applied Physics Express}\ }\textbf {\bibinfo {volume} {10}},\
  \bibinfo {pages} {055503} (\bibinfo {year} {2017})}\BibitemShut {NoStop}%
\bibitem [{\citenamefont {Graupner}\ \emph {et~al.}(1998)\citenamefont
  {Graupner}, \citenamefont {Maier}, \citenamefont {Ristein}, \citenamefont
  {Ley},\ and\ \citenamefont {Jung}}]{RN1492}%
  \BibitemOpen
  \bibfield  {author} {\bibinfo {author} {\bibfnamefont {R.}~\bibnamefont
  {Graupner}}, \bibinfo {author} {\bibfnamefont {F.}~\bibnamefont {Maier}},
  \bibinfo {author} {\bibfnamefont {J.}~\bibnamefont {Ristein}}, \bibinfo
  {author} {\bibfnamefont {L.}~\bibnamefont {Ley}},\ and\ \bibinfo {author}
  {\bibfnamefont {C.}~\bibnamefont {Jung}},\ }\bibfield  {title} {\bibinfo
  {title} {High-resolution surface-sensitive c 1s core-level spectra of clean
  and hydrogen-terminated diamond (100) and (111) surfaces},\ }\href
  {https://doi.org/10.1103/PhysRevB.57.12397} {\bibfield  {journal} {\bibinfo
  {journal} {Physical Review B}\ }\textbf {\bibinfo {volume} {57}},\ \bibinfo
  {pages} {12397} (\bibinfo {year} {1998})}\BibitemShut {NoStop}%
\bibitem [{\citenamefont {Dieckhoff}\ \emph {et~al.}(1999)\citenamefont
  {Dieckhoff}, \citenamefont {Ochs}, \citenamefont {Günster},\ and\
  \citenamefont {Kempter}}]{RN1544}%
  \BibitemOpen
  \bibfield  {author} {\bibinfo {author} {\bibfnamefont {S.}~\bibnamefont
  {Dieckhoff}}, \bibinfo {author} {\bibfnamefont {D.}~\bibnamefont {Ochs}},
  \bibinfo {author} {\bibfnamefont {J.}~\bibnamefont {Günster}},\ and\
  \bibinfo {author} {\bibfnamefont {V.}~\bibnamefont {Kempter}},\ }\bibfield
  {title} {\bibinfo {title} {Metastable impact electron spectroscopy (mies)
  study of chemical vapour deposited (cvd) diamond films},\ }\href
  {https://doi.org/https://doi.org/10.1016/S0039-6028(98)00894-2} {\bibfield
  {journal} {\bibinfo  {journal} {Surface Science}\ }\textbf {\bibinfo {volume}
  {423}},\ \bibinfo {pages} {53} (\bibinfo {year} {1999})}\BibitemShut
  {NoStop}%
\bibitem [{\citenamefont {Ghodbane}\ \emph {et~al.}(2010)\citenamefont
  {Ghodbane}, \citenamefont {Ballutaud}, \citenamefont {Omnès},\ and\
  \citenamefont {Agnès}}]{RN1484}%
  \BibitemOpen
  \bibfield  {author} {\bibinfo {author} {\bibfnamefont {S.}~\bibnamefont
  {Ghodbane}}, \bibinfo {author} {\bibfnamefont {D.}~\bibnamefont {Ballutaud}},
  \bibinfo {author} {\bibfnamefont {F.}~\bibnamefont {Omnès}},\ and\ \bibinfo
  {author} {\bibfnamefont {C.}~\bibnamefont {Agnès}},\ }\bibfield  {title}
  {\bibinfo {title} {Comparison of the xps spectra from homoepitaxial {111},
  {100} and polycrystalline boron-doped diamond films},\ }\href
  {https://doi.org/https://doi.org/10.1016/j.diamond.2010.01.014} {\bibfield
  {journal} {\bibinfo  {journal} {Diamond and Related Materials}\ }\textbf
  {\bibinfo {volume} {19}},\ \bibinfo {pages} {630} (\bibinfo {year}
  {2010})}\BibitemShut {NoStop}%
\bibitem [{\citenamefont {Le~Normand}\ \emph {et~al.}(2001)\citenamefont
  {Le~Normand}, \citenamefont {Hommet}, \citenamefont {Szörényi},
  \citenamefont {Fuchs},\ and\ \citenamefont {Fogarassy}}]{RN1370}%
  \BibitemOpen
  \bibfield  {author} {\bibinfo {author} {\bibfnamefont {F.}~\bibnamefont
  {Le~Normand}}, \bibinfo {author} {\bibfnamefont {J.}~\bibnamefont {Hommet}},
  \bibinfo {author} {\bibfnamefont {T.}~\bibnamefont {Szörényi}}, \bibinfo
  {author} {\bibfnamefont {C.}~\bibnamefont {Fuchs}},\ and\ \bibinfo {author}
  {\bibfnamefont {E.}~\bibnamefont {Fogarassy}},\ }\bibfield  {title} {\bibinfo
  {title} {Xps study of pulsed laser deposited cnx films},\ }\href
  {https://doi.org/10.1103/PhysRevB.64.235416} {\bibfield  {journal} {\bibinfo
  {journal} {Physical Review B}\ }\textbf {\bibinfo {volume} {64}},\ \bibinfo
  {pages} {235416} (\bibinfo {year} {2001})}\BibitemShut {NoStop}%
\bibitem [{\citenamefont {Dementjev}\ \emph {et~al.}(2000)\citenamefont
  {Dementjev}, \citenamefont {de~Graaf}, \citenamefont {van~de Sanden},
  \citenamefont {Maslakov}, \citenamefont {Naumkin},\ and\ \citenamefont
  {Serov}}]{RN1313}%
  \BibitemOpen
  \bibfield  {author} {\bibinfo {author} {\bibfnamefont {A.~P.}\ \bibnamefont
  {Dementjev}}, \bibinfo {author} {\bibfnamefont {A.}~\bibnamefont {de~Graaf}},
  \bibinfo {author} {\bibfnamefont {M.~C.~M.}\ \bibnamefont {van~de Sanden}},
  \bibinfo {author} {\bibfnamefont {K.~I.}\ \bibnamefont {Maslakov}}, \bibinfo
  {author} {\bibfnamefont {A.~V.}\ \bibnamefont {Naumkin}},\ and\ \bibinfo
  {author} {\bibfnamefont {A.~A.}\ \bibnamefont {Serov}},\ }\bibfield  {title}
  {\bibinfo {title} {X-ray photoelectron spectroscopy reference data for
  identification of the c$_{3}$n$_{4}$ phase in carbon–nitrogen films},\
  }\href {https://doi.org/https://doi.org/10.1016/S0925-9635(00)00345-9}
  {\bibfield  {journal} {\bibinfo  {journal} {Diamond and Related Materials}\
  }\textbf {\bibinfo {volume} {9}},\ \bibinfo {pages} {1904} (\bibinfo {year}
  {2000})}\BibitemShut {NoStop}%
\bibitem [{\citenamefont {Ripalda}\ \emph {et~al.}(2000)\citenamefont
  {Ripalda}, \citenamefont {Díaz}, \citenamefont {Román}, \citenamefont
  {Galán}, \citenamefont {Montero}, \citenamefont {Goldoni}, \citenamefont
  {Baraldi}, \citenamefont {Lizzit}, \citenamefont {Comelli},\ and\
  \citenamefont {Paolucci}}]{RN1310}%
  \BibitemOpen
  \bibfield  {author} {\bibinfo {author} {\bibfnamefont {J.~M.}\ \bibnamefont
  {Ripalda}}, \bibinfo {author} {\bibfnamefont {N.}~\bibnamefont {Díaz}},
  \bibinfo {author} {\bibfnamefont {E.}~\bibnamefont {Román}}, \bibinfo
  {author} {\bibfnamefont {L.}~\bibnamefont {Galán}}, \bibinfo {author}
  {\bibfnamefont {I.}~\bibnamefont {Montero}}, \bibinfo {author} {\bibfnamefont
  {A.}~\bibnamefont {Goldoni}}, \bibinfo {author} {\bibfnamefont
  {A.}~\bibnamefont {Baraldi}}, \bibinfo {author} {\bibfnamefont
  {S.}~\bibnamefont {Lizzit}}, \bibinfo {author} {\bibfnamefont
  {G.}~\bibnamefont {Comelli}},\ and\ \bibinfo {author} {\bibfnamefont
  {G.}~\bibnamefont {Paolucci}},\ }\bibfield  {title} {\bibinfo {title}
  {Chemical shift resolved photoionization cross sections of amorphous carbon
  nitride},\ }\href {https://doi.org/10.1103/PhysRevLett.85.2132} {\bibfield
  {journal} {\bibinfo  {journal} {Physical Review Letters}\ }\textbf {\bibinfo
  {volume} {85}},\ \bibinfo {pages} {2132} (\bibinfo {year}
  {2000})}\BibitemShut {NoStop}%
\bibitem [{\citenamefont {Michaelson}\ \emph {et~al.}(2007)\citenamefont
  {Michaelson}, \citenamefont {Lifshitz},\ and\ \citenamefont
  {Hoffman}}]{RN1302}%
  \BibitemOpen
  \bibfield  {author} {\bibinfo {author} {\bibfnamefont {S.}~\bibnamefont
  {Michaelson}}, \bibinfo {author} {\bibfnamefont {Y.}~\bibnamefont
  {Lifshitz}},\ and\ \bibinfo {author} {\bibfnamefont {A.}~\bibnamefont
  {Hoffman}},\ }\bibfield  {title} {\bibinfo {title} {High resolution electron
  energy loss spectroscopy of hydrogenated polycrystalline diamond: Assignment
  of peaks through modifications induced by isotopic exchange},\ }\href
  {https://doi.org/https://doi.org/10.1016/j.diamond.2006.11.079} {\bibfield
  {journal} {\bibinfo  {journal} {Diamond and Related Materials}\ }\textbf
  {\bibinfo {volume} {16}},\ \bibinfo {pages} {855} (\bibinfo {year}
  {2007})}\BibitemShut {NoStop}%
\bibitem [{\citenamefont {Michaelson}\ \emph {et~al.}(2006)\citenamefont
  {Michaelson}, \citenamefont {Hoffman},\ and\ \citenamefont
  {Lifshitz}}]{RN1273}%
  \BibitemOpen
  \bibfield  {author} {\bibinfo {author} {\bibfnamefont {S.}~\bibnamefont
  {Michaelson}}, \bibinfo {author} {\bibfnamefont {A.}~\bibnamefont
  {Hoffman}},\ and\ \bibinfo {author} {\bibfnamefont {Y.}~\bibnamefont
  {Lifshitz}},\ }\bibfield  {title} {\bibinfo {title} {Determination of
  vibrational modes in electron energy loss spectroscopy of polycrystalline
  diamond surfaces by isotopic exchange},\ }\href
  {https://doi.org/10.1063/1.2397027} {\bibfield  {journal} {\bibinfo
  {journal} {Applied Physics Letters}\ }\textbf {\bibinfo {volume} {89}},\
  \bibinfo {pages} {223112} (\bibinfo {year} {2006})}\BibitemShut {NoStop}%
\bibitem [{\citenamefont {Shpilman}\ \emph {et~al.}(2007)\citenamefont
  {Shpilman}, \citenamefont {Gouzman}, \citenamefont {Grossman}, \citenamefont
  {Akhvlediani},\ and\ \citenamefont {Hoffman}}]{RN1315}%
  \BibitemOpen
  \bibfield  {author} {\bibinfo {author} {\bibfnamefont {Z.}~\bibnamefont
  {Shpilman}}, \bibinfo {author} {\bibfnamefont {I.}~\bibnamefont {Gouzman}},
  \bibinfo {author} {\bibfnamefont {E.}~\bibnamefont {Grossman}}, \bibinfo
  {author} {\bibfnamefont {R.}~\bibnamefont {Akhvlediani}},\ and\ \bibinfo
  {author} {\bibfnamefont {A.}~\bibnamefont {Hoffman}},\ }\bibfield  {title}
  {\bibinfo {title} {Oxidation of diamond films by atomic oxygen: High
  resolution electron energy loss spectroscopy studies},\ }\href
  {https://doi.org/10.1063/1.2818373} {\bibfield  {journal} {\bibinfo
  {journal} {Journal of Applied Physics}\ }\textbf {\bibinfo {volume} {102}},\
  \bibinfo {pages} {114914} (\bibinfo {year} {2007})}\BibitemShut {NoStop}%
\bibitem [{\citenamefont {Michaelson}\ \emph {et~al.}(2012)\citenamefont
  {Michaelson}, \citenamefont {Akhvlediani}, \citenamefont {Tkach},\ and\
  \citenamefont {Hoffman}}]{RN1394}%
  \BibitemOpen
  \bibfield  {author} {\bibinfo {author} {\bibfnamefont {S.}~\bibnamefont
  {Michaelson}}, \bibinfo {author} {\bibfnamefont {R.}~\bibnamefont
  {Akhvlediani}}, \bibinfo {author} {\bibfnamefont {L.}~\bibnamefont {Tkach}},\
  and\ \bibinfo {author} {\bibfnamefont {A.}~\bibnamefont {Hoffman}},\
  }\bibfield  {title} {\bibinfo {title} {Evidence for preferential reactivity
  of the atomic oxygen with hydrogenated diamond (111) facets},\ }\href
  {https://doi.org/https://doi.org/10.1016/j.susc.2012.05.008} {\bibfield
  {journal} {\bibinfo  {journal} {Surface Science}\ }\textbf {\bibinfo {volume}
  {606}},\ \bibinfo {pages} {L79} (\bibinfo {year} {2012})}\BibitemShut
  {NoStop}%
\bibitem [{\citenamefont {Attrash}\ \emph {et~al.}(2020)\citenamefont
  {Attrash}, \citenamefont {Kuntumalla}, \citenamefont {Michaelson},\ and\
  \citenamefont {Hoffman}}]{RN1361}%
  \BibitemOpen
  \bibfield  {author} {\bibinfo {author} {\bibfnamefont {M.}~\bibnamefont
  {Attrash}}, \bibinfo {author} {\bibfnamefont {M.~K.}\ \bibnamefont
  {Kuntumalla}}, \bibinfo {author} {\bibfnamefont {S.}~\bibnamefont
  {Michaelson}},\ and\ \bibinfo {author} {\bibfnamefont {A.}~\bibnamefont
  {Hoffman}},\ }\bibfield  {title} {\bibinfo {title} {Nitrogen-terminated
  polycrystalline diamond surfaces by microwave chemical vapor deposition:
  Thermal stability, chemical states, and electronic structure},\ }\href
  {https://doi.org/10.1021/acs.jpcc.9b10829} {\bibfield  {journal} {\bibinfo
  {journal} {Journal of Physical Chemistry C}\ }\textbf {\bibinfo {volume}
  {124}},\ \bibinfo {pages} {5657} (\bibinfo {year} {2020})}\BibitemShut
  {NoStop}%
\bibitem [{\citenamefont {Yamano}\ \emph {et~al.}(2017)\citenamefont {Yamano},
  \citenamefont {Kawai}, \citenamefont {Kato}, \citenamefont {Kageura},
  \citenamefont {Inaba}, \citenamefont {Okada}, \citenamefont {Higashimata},
  \citenamefont {Haruyama}, \citenamefont {Tanii}, \citenamefont {Yamada} \emph
  {et~al.}}]{yamano2017charge}%
  \BibitemOpen
  \bibfield  {author} {\bibinfo {author} {\bibfnamefont {H.}~\bibnamefont
  {Yamano}}, \bibinfo {author} {\bibfnamefont {S.}~\bibnamefont {Kawai}},
  \bibinfo {author} {\bibfnamefont {K.}~\bibnamefont {Kato}}, \bibinfo {author}
  {\bibfnamefont {T.}~\bibnamefont {Kageura}}, \bibinfo {author} {\bibfnamefont
  {M.}~\bibnamefont {Inaba}}, \bibinfo {author} {\bibfnamefont
  {T.}~\bibnamefont {Okada}}, \bibinfo {author} {\bibfnamefont
  {I.}~\bibnamefont {Higashimata}}, \bibinfo {author} {\bibfnamefont
  {M.}~\bibnamefont {Haruyama}}, \bibinfo {author} {\bibfnamefont
  {T.}~\bibnamefont {Tanii}}, \bibinfo {author} {\bibfnamefont
  {K.}~\bibnamefont {Yamada}}, \emph {et~al.},\ }\bibfield  {title} {\bibinfo
  {title} {Charge state stabilization of shallow nitrogen vacancy centers in
  diamond by oxygen surface modification},\ }\href
  {https://doi.org/10.7567/jjap.56.04ck08} {\bibfield  {journal} {\bibinfo
  {journal} {Japanese Journal of Applied Physics}\ }\textbf {\bibinfo {volume}
  {56}},\ \bibinfo {pages} {04CK08} (\bibinfo {year} {2017})}\BibitemShut
  {NoStop}%
\bibitem [{\citenamefont {de~Sousa}(2009)}]{deSousa2009}%
  \BibitemOpen
  \bibfield  {author} {\bibinfo {author} {\bibfnamefont {R.}~\bibnamefont
  {de~Sousa}},\ }\bibinfo {title} {Electron spin as a spectrometer
  of nuclear-spin noise and other fluctuations},\ in\ \href
  {https://doi.org/10.1007/978-3-540-79365-6_10} {\emph {\bibinfo {booktitle}
  {Electron Spin Resonance and Related Phenomena in Low-Dimensional
  Structures}}},\ \bibinfo {editor} {edited by\ \bibinfo {editor}
  {\bibfnamefont {M.}~\bibnamefont {Fanciulli}}}\ (\bibinfo  {publisher}
  {Springer Berlin Heidelberg},\ \bibinfo {address} {Berlin, Heidelberg},\
  \bibinfo {year} {2009})\ pp.\ \bibinfo {pages} {183--220}\BibitemShut
  {NoStop}%
\bibitem [{\citenamefont {Stanwix}\ \emph {et~al.}(2010)\citenamefont
  {Stanwix}, \citenamefont {Pham}, \citenamefont {Maze}, \citenamefont
  {Le~Sage}, \citenamefont {Yeung}, \citenamefont {Cappellaro}, \citenamefont
  {Hemmer}, \citenamefont {Yacoby}, \citenamefont {Lukin},\ and\ \citenamefont
  {Walsworth}}]{stanwix2010exponent}%
  \BibitemOpen
  \bibfield  {author} {\bibinfo {author} {\bibfnamefont {P.~L.}\ \bibnamefont
  {Stanwix}}, \bibinfo {author} {\bibfnamefont {L.~M.}\ \bibnamefont {Pham}},
  \bibinfo {author} {\bibfnamefont {J.~R.}\ \bibnamefont {Maze}}, \bibinfo
  {author} {\bibfnamefont {D.}~\bibnamefont {Le~Sage}}, \bibinfo {author}
  {\bibfnamefont {T.~K.}\ \bibnamefont {Yeung}}, \bibinfo {author}
  {\bibfnamefont {P.}~\bibnamefont {Cappellaro}}, \bibinfo {author}
  {\bibfnamefont {P.~R.}\ \bibnamefont {Hemmer}}, \bibinfo {author}
  {\bibfnamefont {A.}~\bibnamefont {Yacoby}}, \bibinfo {author} {\bibfnamefont
  {M.~D.}\ \bibnamefont {Lukin}},\ and\ \bibinfo {author} {\bibfnamefont
  {R.~L.}\ \bibnamefont {Walsworth}},\ }\bibfield  {title} {\bibinfo {title}
  {Coherence of nitrogen-vacancy electronic spin ensembles in diamond},\ }\href
  {https://doi.org/10.1103/PhysRevB.82.201201} {\bibfield  {journal} {\bibinfo
  {journal} {Physical Review B}\ }\textbf {\bibinfo {volume} {82}},\ \bibinfo
  {pages} {201201} (\bibinfo {year} {2010})}\BibitemShut {NoStop}%
\bibitem [{\citenamefont {Attrash}\ \emph {et~al.}(2019)\citenamefont
  {Attrash}, \citenamefont {Kuntumalla},\ and\ \citenamefont
  {Hoffman}}]{RN1362}%
  \BibitemOpen
  \bibfield  {author} {\bibinfo {author} {\bibfnamefont {M.}~\bibnamefont
  {Attrash}}, \bibinfo {author} {\bibfnamefont {M.~K.}\ \bibnamefont
  {Kuntumalla}},\ and\ \bibinfo {author} {\bibfnamefont {A.}~\bibnamefont
  {Hoffman}},\ }\bibfield  {title} {\bibinfo {title} {Bonding, structural
  properties and thermal stability of low damage rf (n2) plasma treated diamond
  (100) surfaces studied by xps, leed, and tpd},\ }\href
  {https://doi.org/https://doi.org/10.1016/j.susc.2018.11.006} {\bibfield
  {journal} {\bibinfo  {journal} {Surface Science}\ }\textbf {\bibinfo {volume}
  {681}},\ \bibinfo {pages} {95} (\bibinfo {year} {2019})}\BibitemShut
  {NoStop}%
\bibitem [{\citenamefont {Shirley}(1972)}]{RN1489}%
  \BibitemOpen
  \bibfield  {author} {\bibinfo {author} {\bibfnamefont {D.~A.}\ \bibnamefont
  {Shirley}},\ }\bibfield  {title} {\bibinfo {title} {High-resolution x-ray
  photoemission spectrum of the valence bands of gold},\ }\href
  {https://doi.org/10.1103/PhysRevB.5.4709} {\bibfield  {journal} {\bibinfo
  {journal} {Physical Review B}\ }\textbf {\bibinfo {volume} {5}},\ \bibinfo
  {pages} {4709} (\bibinfo {year} {1972})}\BibitemShut {NoStop}%
\bibitem [{\citenamefont {Biesinger}\ \emph {et~al.}(2011)\citenamefont
  {Biesinger}, \citenamefont {Payne}, \citenamefont {Grosvenor}, \citenamefont
  {Lau}, \citenamefont {Gerson},\ and\ \citenamefont {Smart}}]{RN1490}%
  \BibitemOpen
  \bibfield  {author} {\bibinfo {author} {\bibfnamefont {M.~C.}\ \bibnamefont
  {Biesinger}}, \bibinfo {author} {\bibfnamefont {B.~P.}\ \bibnamefont
  {Payne}}, \bibinfo {author} {\bibfnamefont {A.~P.}\ \bibnamefont
  {Grosvenor}}, \bibinfo {author} {\bibfnamefont {L.~W.~M.}\ \bibnamefont
  {Lau}}, \bibinfo {author} {\bibfnamefont {A.~R.}\ \bibnamefont {Gerson}},\
  and\ \bibinfo {author} {\bibfnamefont {R.~S.~C.}\ \bibnamefont {Smart}},\
  }\bibfield  {title} {\bibinfo {title} {Resolving surface chemical states in
  xps analysis of first row transition metals, oxides and hydroxides: Cr, mn,
  fe, co and ni},\ }\href
  {https://doi.org/https://doi.org/10.1016/j.apsusc.2010.10.051} {\bibfield
  {journal} {\bibinfo  {journal} {Applied Surface Science}\ }\textbf {\bibinfo
  {volume} {257}},\ \bibinfo {pages} {2717} (\bibinfo {year}
  {2011})}\BibitemShut {NoStop}%
\bibitem [{\citenamefont {Biesinger}\ \emph {et~al.}(2010)\citenamefont
  {Biesinger}, \citenamefont {Lau}, \citenamefont {Gerson},\ and\ \citenamefont
  {Smart}}]{RN1491}%
  \BibitemOpen
  \bibfield  {author} {\bibinfo {author} {\bibfnamefont {M.~C.}\ \bibnamefont
  {Biesinger}}, \bibinfo {author} {\bibfnamefont {L.~W.~M.}\ \bibnamefont
  {Lau}}, \bibinfo {author} {\bibfnamefont {A.~R.}\ \bibnamefont {Gerson}},\
  and\ \bibinfo {author} {\bibfnamefont {R.~S.~C.}\ \bibnamefont {Smart}},\
  }\bibfield  {title} {\bibinfo {title} {Resolving surface chemical states in
  xps analysis of first row transition metals, oxides and hydroxides: Sc, ti,
  v, cu and zn},\ }\href
  {https://doi.org/https://doi.org/10.1016/j.apsusc.2010.07.086} {\bibfield
  {journal} {\bibinfo  {journal} {Applied Surface Science}\ }\textbf {\bibinfo
  {volume} {257}},\ \bibinfo {pages} {887} (\bibinfo {year}
  {2010})}\BibitemShut {NoStop}%
\bibitem [{\citenamefont {Pham}(2013)}]{pham2013thesis}%
  \BibitemOpen
  \bibfield  {author} {\bibinfo {author} {\bibfnamefont {L.~M.}\ \bibnamefont
  {Pham}},\ }\emph {\bibinfo {title} {Magnetic Field Sensing with
  Nitrogen-Vacancy Color Centers in Diamond}},\ \href
  {http://nrs.harvard.edu/urn-3:HUL.InstRepos:11051173} {Ph.D. thesis},\
  \bibinfo  {school} {Harvard University} (\bibinfo {year} {2013})\BibitemShut
  {NoStop}%
\end{thebibliography}%

\end{document}